\documentclass[superscriptadress,A4,twocolumn,10pt,showpacs]{revtex4-1}
\usepackage{epstopdf}
\usepackage{graphicx}
\usepackage{amsmath}
\usepackage{amsfonts}
\usepackage{amssymb}
\usepackage{xspace}
\usepackage{verbatim}
\usepackage[francais]{babel}
\usepackage[latin1]{inputenc}
\usepackage{rotating}

\newcommand{\EVRY}{D\'epartement de Physique,
Universit\'e d'Evry Val d'Essonne\\
Rue du p\`ere Andr\'e Jarlan, 91025 Evry cedex}
\newcommand{\LKB}{Laboratoire Kastler Brossel UMR 8552, UPMC, CNRS, ENS, C 74, Universit\'e Pierre et Marie Curie\\
4 place Jussieu, 75252 Paris, France}
\newcommand{\Amsterdam}{LaserLaB, VU University, De Boelelaan 1081,
1081 HV Amsterdam, The Netherlands}

\begin{document}

\title{Two-photon spectroscopy of trapped HD$^+$ ions in the Lamb-Dicke regime}
\author{Vu Quang Tran }
\affiliation{\LKB}
\author{Jean-Philippe Karr}
\affiliation{\LKB}
\affiliation{\EVRY}
\author{Albane Douillet}
\affiliation{\LKB}
\affiliation{\EVRY}
\author{Jeroen C. J. Koelemeij}
\affiliation{\Amsterdam}
\author{Laurent Hilico}
\affiliation{\LKB}
\affiliation{\EVRY}
% \date\today
\begin{abstract}
We study the feasibility of nearly-degenerate two-photon rovibrational spectroscopy in ensembles of trapped, sympathetically cooled hydrogen molecular ions using a resonance-enhanced multiphoton dissociation (REMPD) scheme. Taking advantage of quasi-coincidences in the rovibrational spectrum, the excitation lasers are tuned close to an intermediate level to resonantly enhance two-photon absorption. Realistic simulations of the REMPD signal are obtained using a four-level model that takes into account saturation effects, ion trajectories, laser frequency noise and redistribution of population by blackbody radiation. We show that the use of counterpropagating laser beams enables optical excitation in an effective Lamb-Dicke regime. Sub-Doppler lines having widths in the 100 Hz range can be observed with good signal-to-noise ratio for an optimal choice of laser detunings. Our results indicate the feasibility of molecular spectroscopy at the $10^{-14}$ accuracy level for improved tests of molecular QED, a new determination of the proton-to-electron mass ratio, and studies of the time (in)dependence of the latter.
\end{abstract}
\maketitle

\section{Introduction}

Most high-precision measurements in atomic or molecular physics rely on laser spectroscopy in dilute gases. Several methods have been developed to suppress Doppler line broadening and reach natural linewidth or laser width limited resolutions, such as saturated absorption~\cite{Borde1970,Smith1971,Hansch1971} or Doppler-free two-photon spectroscopy~\cite{Vasilenko1970,Cagnac1973,Biraben1974,Levenson1974}. Resonantly enhanced two-photon absorption using two lasers of unequal frequencies tuned close to an intermediate level was also studied both theoretically and experimentally~\cite{Bjorkholm1974,Bjorkholm1976,Grove1995}; sequential two-photon absorption at exact resonance was shown to provide both maximum transition rates and Doppler-free spectra. Indeed, the photon absorbed in the first transition selects a velocity class from which the second absorption occurs without Doppler broadening.

One of the most successful methods to suppress the Doppler effect is single-photon absorption on trapped species in the Lamb-Dicke regime where the confinement length is smaller than the wavelength. This condition is easily satisfied in ion traps in the microwave domain, which has allowed high-precision hyperfine structure measurements in many ionic species~\cite{Fortson1966,Richardson1968,Werth1995} and the development of microwave frequency standards~\cite{Fisk1997,Burt2008,Jau2012}. The Lamb-Dicke regime is much more challenging to achieve in the optical domain~\cite{Neuhauser1978,Bergquist1987}. It requires tight confinement of laser-cooled ions, and has only been obtained with small ion numbers, i.e. single ions or ion strings located on the axis of a linear trap.

We address here the specific case of molecular ions, where high-resolution infrared spectroscopy opens the way to many interesting applications such as tests of QED~\cite{Korobov2009,Korobov2013} or parity violation~\cite{Borschevsky2012}, measurement of nucleus-to-electron mass ratios~\cite{Wing1976,Roth2008}, and studies of their variation in time~\cite{Schiller2005,Flambaum2007,Shelkovnikov2008}. Studies on small ion numbers in the Lamb-Dicke regime raise additional problems due to the difficulty of preparing and controlling the internal state of molecules. So far, the best resolutions have been obtained with ensembles of sympathetically cooled molecular ions~\cite{Koelemeij2007,Bressel2012}. Temperatures of a few tens of mK are typically achieved, which corresponds to a Doppler broadening of several MHz, well above the natural linewidths of excited rovibrational states.

To circumvent this limitation, degenerate Doppler-free two-photon spectroscopy is a natural solution~\cite{Hilico2001,Karr2005}. However, relatively high field intensities are generally required to achieve a substantial transition rate, and this approach often implies installing a high-finesse enhancement cavity in the vacuum chamber~\cite{Karr2008}. For the sake of experimental convenience and universality, a sub-Doppler spectroscopic scheme that would be free of this requirement is highly desirable.

In this paper, we analyze theoretically the resonantly enhanced two-photon excitation of trapped molecular ions with nearly-degenerate counterpropagating laser fields, that is made possible by quasi-coincidences in the rovibrational spectrum. Near-resonant excitation of an intermediate level warrants sufficient transition rates with moderate laser power; in addition, two-photon absorption takes place in the Lamb-Dicke regime, due to the effective wavelength associated with simultaneous absorption of one photon from each field. The proposed scheme thus combines advantages of the resonant enhancement already evidenced in neutral gases, and of the Lamb-Dicke effect that has been exploited in microwave spectroscopy of trapped ions.

As a first application, we will focus on hydrogen molecular ions. These simple systems enable highly precise comparisons between measured transition frequencies and theoretical predictions. Current efforts to evaluate hyperfine structure~\cite{Korobov2009} and QED corrections~\cite{Korobov2013} in H$_2^+$ or HD$^+$ are expected to improve the theoretical accuracy beyond 0.1~ppb, allowing for stringent tests of QED, and for an improved determination of the proton-to-electron mass ratio (presently known to 0.41~ppb accuracy~\cite{Mohr2010}). The high Q-factor of rovibrational lines also opens the way to searches for possible time variations of fundamental constants~\cite{Schiller2005,Flambaum2007} and 'fifth forces'~\cite{Salumbides2013} with improved sensitivity. Experimental studies on sympathetically cooled HD$^+$ ions~\cite{Bressel2012,Koelemeij2012}, using single-photon rovibrational transitions detected by (1+1') resonance-enhanced multiphoton dissociation (REMPD), are so far limited to the ppb level, mainly by the Doppler broadening. As we will show, this limitation can be overcome by several orders of magnitude in the proposed experiment.

The paper is organized as follows. In Sec.~\ref{sec_time_freq_scale}, we describe the proposed (1+1'+1'') REMPD experimental scheme, and discuss all the frequency scales in the problem. The theoretical model of REMPD is introduced in~Sec.~\ref{sec_model}. We model the molecule-light interaction as a three-level system which interacts coherently with the two laser fields, and take dissociation into account by introducing a non-coherent coupling to a fourth level. Our treatment furthermore includes the motional degrees of freedom of the molecules. The dynamics of the entire system are captured within a set of optical Bloch equations (OBE), which are solved to predict the dissociated ion fraction monitored in the experiment. In Sec.~\ref{sec_results}, we first numerically solve the model in the ideal case of a single ion undergoing a pure harmonic motion in order to highlight the main features of the signal and evidence the Lamb-Dicke effect. Realistic ion motion obtained from molecular ion dynamics simulations is then incorporated in the model, and optimal conditions for the experiment in terms of laser detunings, which are found markedly different from the gas case~\cite{Bjorkholm1976}, are determined. We show that under these conditions, an approximate model of the two-photon transition rate can be used, and its validity range is assessed by comparing to the exact OBE model. The power shift and broadening is analyzed, as well as the effect of laser frequency noise. Finally, in Sec.~\ref{sec_BBR}, in order to obtain realistic estimates of the expected REMPD signal strength, we simulate the dynamics of the total number of HD$^+$ ions taking into account the REMPD rates as well as the redistribution of rotational population induced by black-body radiation (BBR).

\section{Two-photon transitions in HD$^+$ and frequency scales of the problem}\label{sec_time_freq_scale}

The permanent electric dipole moment of HD$^+$ allows rovibrational transitions within the electronic ground state. Weak vibrational overtone transitions exist only by virtue of the anharmonicity of the HD$^+$ bond. Two-photon vibrational transitions are possible, but require a quasi-resonance with an intermediate level to achieve sufficiently high transition rates. Using the extensive set of accurate rovibrational level energies obtained by Moss~\cite{Moss1993}, an analysis of intermediate level energy mismatch reveals two interesting transitions, $(v\!=\!0,L\!=\!1)\!\rightarrow\!(v\!=\!1,L\!=\!0)\!\rightarrow\!(v\!=\!2,L\!=\!1)$
at 5.37~$\mu$m~\cite{Karr2005} (energy mismatch: $\Delta E = 6.18$~cm$^{-1}$) and $(v\!=\!0,L\!=\!3)\!\rightarrow\!(v\!=\!4,L\!=\!2)\!\rightarrow\!(v\!=\!9,L\!=\!3)$ at 1.44~$\mu$m ($\Delta E = 6.84$~cm$^{-1}$). In the following, we will consider the latter, whose wavelength is more convenient for laser stabilization and absolute frequency measurements.

Throughout the paper, the values of various parameters are taken from the HD$^+$ spectroscopy experiment developed by the Amsterdam team and described in~\cite{Koelemeij2012}. A set of about 100 HD$^+$ ions is sympathetically cooled by 1-2~10$^3$ laser-cooled Be$^+$ ions to about 10~mK. The (1+1'+1") REMPD experiment proposed here consists in driving a quasi-degenerate two-photon overtone transition using counterpropagating beams. The $v=9$ level is efficiently photodissociated using a 532~nm laser beam. Two-photon excitation and subsequent dissociation lead to loss of HD$^+$ ions from the trapped ensemble. This loss is observed by comparing the number of HD$^+$ ions before and after REMPD, which is deduced from the fluorescence photons emitted by the laser-cooled Be$^+$ ions while heating the ion ensemble through  resonant excitation of the HD$^+$ motion~\cite{Roth2006}. The detection noise typically observed in the experiment limits the minimum detectable  dissociated HD$^+$ fraction to a few percent.

\begin{figure}%[!h]
  \includegraphics[width=8cm]{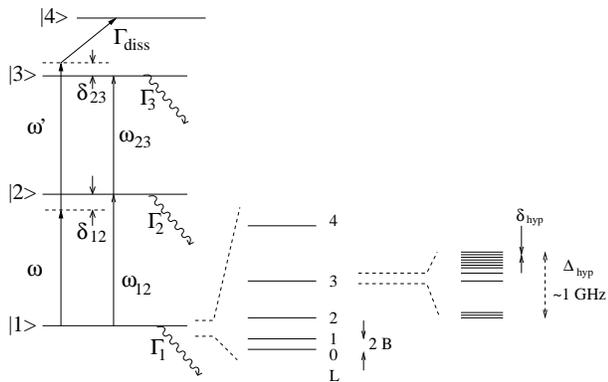}\\
  \caption{Sketch of the HD$^+$ energy levels involved in the proposed REMPD experiment. Left: vibrational structure; couplings by laser fields and spontaneous relaxation are respectively indicated by straight and zigzag arrows. Center: rotational structure. Right: hyperfine structures (not to scale).}
  \label{fig1_HDlevels}
\end{figure}

Figure~\ref{fig1_HDlevels} shows the structure of HD$^+$ energy levels involved in the REMPD scheme. The kets $|1\rangle$, $|2\rangle$ and $|3\rangle$ denote the levels $(v\!=\!0,L\!=\!3,F,S,J_1)$, $(v\!=\!4,L\!=\!2,F,S,J_2)$ and $(v\!=\!9,L\!=\!3,F,S,J_3)$, where $(F,S,J)$ are hyperfine quantum numbers according to the coupling scheme detailed below. The vibrational structure (with intervals of about 60~THz) sets the larger frequency scale in the experiment, followed by the rotational constant ($B\!\approx 700$~GHz). The resonant angular frequencies are $\omega_{12}/2\pi=207.838$~THz and $\omega_{23}/2\pi=207.427$~THz leading to a small two-photon transition mismatch $\omega_{12}\!\!-\!\!\omega_{23}\!\!=410$~GHz (13.7~cm$^{-1}$), i.e. 0.2\% in relative value. The rovibrational states have small natural widths $\Gamma_1/2\pi=0.037$~Hz, $\Gamma_2/2\pi=9.2$~Hz and $\Gamma_3/2\pi=13.1$~Hz~\cite{Amitay1994}.

We use the standard spin coupling scheme $\mathbf{F}=\mathbf{I_p}+\mathbf{S_e}$, $\mathbf{S}=\mathbf{F}+\mathbf{I_d}$, $\mathbf{J}=\mathbf{S}+\mathbf{L}$~\cite{Bakalov2006}. For a given rovibrational level, the hyperfine structure spreads over $\Delta_{rm hyp}\approx 1$~GHz, the smallest interval between two hyperfine sub-levels being $\delta_{rm hyp}\approx 8$~MHz for the v=0, L=3 level~\cite{Bakalov2006}. The Zeeman structure is discussed in Appendix~\ref{ap_Zeeman effect}. In the Amsterdam experimental setup, the magnetic field can be reduced to values as low as 20~mG, resulting in a Zeeman splitting $\delta_Z$ smaller than 10~Hz for the best suited lines, due to an almost perfect compensation of Zeeman shifts. Assuming the laser linewidth is larger than the Zeeman splitting, all the Zeeman components can be addressed simultaneously. That is why the Zeeman structure is not considered in our model.

The two-photon transition is driven by two lasers of angular frequencies $\omega$ and $\omega'$ close to the single-photon transition frequencies $\omega_{12}$ and $\omega_{23}$ respectively. The transitions lye in the 1.4~$\mu$m range and can be driven by frequency-stabilized diode lasers. Using transition moments calculated with the approach of~\cite{Koelemeij2011}, the achievable laser intensities i.e. about 20~mW/(0.1 mm)$^2$ can be translated into maximum values for the Rabi frequencies $\Omega_{12}/2\pi$ or $\Omega_{23}/2\pi$ that exceed 100~kHz.

In the linear RF Paul trap, the HD$^+$ molecular ions are embedded in a Be$^+$ Coulomb crystal~\cite{Drewsen2003,Blythe2005}. As part of a complex mechanical system, each HD$^+$ ion oscillates around its equilibrium position with oscillation frequencies in the kHz to MHz ranges and amplitudes in the $\mu$m range. Velocities can typically reach 5~m/s (see Appendix~\ref{ap_ion_dynamics}) resulting in a single-photon Doppler effect $\Gamma_{D}$ in the 5~MHz range that clearly dominates the single-photon transition width. The values of the laser detunings $\delta_{12}=\omega-\omega_{12}$ and $\delta_{23}=\omega'-\omega_{23}$ with respect to $\Gamma_{D}$ determine the dynamics of the system: instantaneous two-photon transitions will dominate if $\delta_{12} > \Gamma_{D}$, while in the opposite case sequential transitions will also take place.

With a counterpropagating two-photon excitation scheme, the effective wavelength $\lambda_{\rm eff}=2\pi c/(|\omega_{12}-\omega_{23}|)$ is about 500 times larger than the single-photon wavelength, i.e. 0.7~mm. Since the ion motional amplitude $a$ is about 1~$\mu$m, the Lamb-Dicke parameter $\eta=a/\lambda_{\rm eff}\approx$~0.014 is much smaller than unity, leading to a Doppler-free signal, as will be evidenced in Sec.~\ref{sec_results}.

The REMPD process involves a photodissociation step from level $|3\rangle$ using a 532~nm cw laser with a maximum intensity of 140~W/cm$^2$, corresponding to a photodissociation rate $\Gamma_{\rm diss}=5000$~s$^{-1}$. In the following, we use  $\Gamma_{\rm diss}=200$~s$^{-1}$, which is still much larger than the natural decay rate, and sufficient to detect the REMPD signal~\cite{Koelemeij2012}.

The laser linewidths $\Gamma_{L}$ may range from hundreds of kHz down to the Hz level depending on the laser frequency stabilization scheme. In case of imperfect stabilization, $\Gamma_{L}$ may be comparable to the Rabi frequencies and strongly affect the two-photon transition rate and linewidth, which requires taking laser frequency noise into account.

At thermal equilibrium at room temperature, most of the HD$^+$ population is concentrated in the $v=0$, $0\leq L \leq 5$ levels~\cite{Koelemeij2007b}. Blackbody radiation permanently redistributes the populations among those levels with transition rates $\Gamma_{\rm BBR}$ in the 0.1~s$^{-1}$ range\cite{Koelemeij2011}, the smallest frequency scale of the problem.

To summarize, the different rates follow the hierarchy
\begin{eqnarray}
 \nonumber \Gamma_{\rm BBR}\ll\delta_Z,\Gamma_{1,2,3}\leq\Gamma_{\rm diss},\Gamma_L\leq\Omega_{12}\approx\Omega_{23} \\
           \ll\Gamma_D\sim\delta_{12}\sim\delta_{23}\leq\delta_{\rm hyp}\\
 \nonumber \ll\Delta_{\rm hyp}\ll|\omega_{12}-\omega_{23}|<B\ll \omega_{12}, \omega_{23}.
\end{eqnarray}
This analysis shows that the different hyperfine components of the two-photon transition can be considered to be well isolated, and that it is appropriate to study the two-photon transition rate using a three-level ladder system. It also shows that one can distinguish two different time scales for the population evolution: a fast one due to laser couplings and spontaneous relaxation, and a much slower one due to BBR population redistribution. As a consequence, in a very good approximation, the REMPD process can be studied in two steps. The first step evaluates the short-term ($\approx$~1s) time evolution of a three-level system under laser excitation and spontaneous decay to obtain the effective two-photon excitation and REMPD rates (Secs.~\ref{sec_model} and \ref{sec_results}). In the second step, the long-term evolution of the total number of HD$^+$ ions is studied, taking into account the REMPD rate obtained in the first step, and the redistribution of rotational-state population by BBR (Sec.~\ref{sec_BBR}).

\section{REMPD model}\label{sec_model}

We consider the three-level ladder structure shown in Fig.~\ref{fig1_HDlevels}. For states $|2\rangle$ and $|3\rangle$, the relaxation by spontaneous emission mainly populates rovibrational levels with $v'=v-1$. The spontaneous emission cascade, coupled to BBR reditribution, can of course ultimately populate the $v\!=\!0, L\!=\!3$ state, but this happens on much longer time scales with respect to laser excitation, dissociation and spontaneous decay. We thus treat the three-level system as an open system, and postpone the analysis of BBR redistribution to Sec.~\ref{sec_BBR}. While levels $|1\rangle$ and $|2\rangle$ have natural widths $\Gamma_1$ and $\Gamma_2$, level $|3\rangle$ relaxes through spontaneous emission with a natural width $\Gamma_3$ and through dissociation with a rate $\Gamma_{\rm diss}$ resulting in an effective width $\Gamma_3^{\rm eff}=\Gamma_3+\Gamma_{\rm diss}$. We introduce a fourth virtual level $|4\rangle$ whose population represents the photodissociated fraction. The coupling to level $|4\rangle$ is an irreversible process.

The ions are excited by two counterpropagating beams of angular frequencies $\omega$ and $\omega'$ close to the resonant frequencies $\omega_{12}$ and $\omega_{23}$. The corresponding electric field is given by:
\begin{equation}\label{eq_E_field}
{\bf E}({\bf r},t)=E\boldsymbol{\epsilon}e^{-i(\omega t-{\bf k}.{\bf r}+\varphi(t))}+E\boldsymbol{\epsilon}'e^{-i(\omega' t-{\bf k}'.{\bf r}+\varphi'(t))}+c.c.
\end{equation}
where $\varphi(t)$ and $\varphi'(t)$ describe laser phase noise, and $E, E'$ and $\boldsymbol{\epsilon,\epsilon}'$ stand for the field amplitudes and polarization states, respectively.

Following the lines of~\cite{Hansch1970}, the density matrix $\varrho({\bf r},t)$ obeys the optical Bloch equations (OBE) $\frac{d}{dt}\varrho({\bf r},t)=\frac{1}{i\hbar}[H,\varrho({\bf r},t)]+\dot{\varrho}_{\rm relax}$ where the total time derivative is written as $\frac{d}{dt}=\frac{\partial}{\partial t}+{\bf v}.{\bf \nabla}$. Applying the rotating wave approximation, we set
\begin{equation}\label{eq_rot_wave}
\left\{
\begin{array}{ccl}
\varrho_{ii}&=&\rho_{ii},\ i=1..4\\
\varrho_{12}&=&\rho_{12}(t)e^{-i(\omega t-{\bf k}.{\bf r}(t))} \\
\varrho_{23}&=&\rho_{23}(t)e^{-i(\omega' t-{\bf k}'.{\bf r}(t))} \\
\varrho_{13}&=&\rho_{13}(t)e^{-i((\omega+\omega') t-({\bf k}+{\bf k}').{\bf r}(t))}
\end{array}
\right.
\end{equation}
and we obtain
\begin{eqnarray}
 \dot{\rho_{11}} &=& -\Gamma_{1}\rho_{11} \nonumber
     +i\left(\Omega_{12}\rho_{21}-\Omega_{12}^*\rho_{12} \right) \\ \nonumber
\dot{\rho_{22}} &=& -\Gamma_{2}\rho_{22}
      +i\left(\Omega_{12}^*\rho_{12}-\Omega_{12}\rho_{21}+\Omega_{23}\rho_{32}-\Omega_{23}^*\rho_{23} \right) \\ \nonumber
\dot{\rho_{33}} &=& -\left(\Gamma_{3}+\Gamma_{\rm diss}\right)\rho_{33}
     +i\left(\Omega_{23}^*\rho_{23}-\Omega_{23}\rho_{32} \right) \\ \nonumber
\dot{\rho_{44}} &=& \Gamma_{\rm diss}\rho_{33} \\ \nonumber
\dot{\rho_{12}} &=& \left(i(\delta_{12}-{\bf k}.{\bf \dot{r}}(t))-\gamma_{12}\right)\rho_{12}\\ \label{eq_OBE}
                & & +i\left(\Omega_{12}(\rho_{22}-\rho_{11})-\Omega_{23}^*\rho_{13} \right)\\ \nonumber
\dot{\rho_{13}} &=& \left(i(\delta_{12}+\delta_{23}-({\bf k}+{\bf k}').{\bf \dot{r}}(t))-\gamma_{13}\right)\rho_{13}\\ \nonumber
                & & +i\left(\Omega_{12}\rho_{23}-\Omega_{23}\rho_{12} \right)\\ \nonumber
\dot{\rho_{23}} &=& \left(i(\delta_{23}-{\bf k}'.{\bf \dot{r}}(t))-\gamma_{23}\right)\rho_{23}\\ \nonumber
                & & +i\left(\Omega_{23}(\rho_{33}-\rho_{22})+\Omega_{12}^*\rho_{13} \right) \nonumber
\end{eqnarray}
where the Rabi frequencies are $\Omega_{12}={\bf d_{12}}.\boldsymbol{\epsilon}Ee^{i\varphi(t)}/\hbar$ and $\Omega_{23}={\bf d_{23}}.\boldsymbol{\epsilon}'E'e^{i\varphi'(t)}/\hbar$ with the dipole moment matrix elements ${\bf d}_{ij}=\langle i|{\bf d}|j\rangle$. The coherences relaxation rates are $\gamma_{12}\!\!=\!\!(\Gamma_1+\Gamma_2)/2$ and $\gamma_{i3}\!\!=\!\!(\Gamma_i+\Gamma_3+\Gamma_{\rm diss})/2$.
The photodissociated fraction $\rho_{44}$ is proportional to the time-integral of the upper level population $\rho_{33}(t)$.

The Doppler effect appears in the terms ${\bf k}.{\bf \dot{r}}(t)$ and ${\bf k'}.{\bf \dot{r}}(t)$ in the evolution equations of $\rho_{12}$ and $\rho_{23}$. Suppression of the Doppler effect occurs in the $\rho_{13}$ evolution equation in the case of counterpropagating beams of nearly equal frequencies
for which ${\bf k}+{\bf k}'\approx {\bf 0}$. In the following, the laser direction is assumed parallel to the linear trap axis $z$, so that the Doppler effect
in Eq.~(\ref{eq_OBE}) reduces to ${\bf k}.{\dot{\bf r}}(t)=k\ {\dot z}(t)$ and ${\bf k}'.{\dot{\bf r}}(t)=-k'\ {\dot z}(t)$. This assumption furthermore justifies ignoring effects of ion micromotion at the RF trap frequency.
A detailed discussion of micromotion effects is postponed to Sec.~\ref{sec_micromotion}.

At first glance, the largest REMPD signal could be expected for the doubly resonant configuration, $\delta_{12}=-\delta_{23}=0$, as in a thermal gas~\cite{Bjorkholm1976}. However, if the detunings are smaller than (or comparable to) the single-photon Doppler width, sequential absorption of photons $\omega$ and $\omega'$ through level $|2\rangle$ can compete with the Doppler-free signal. The main objective of this paper is to determine the experimental conditions under which one can obtain sub-Doppler REMPD signals with the largest signal-to-noise ratio; in particular, to determine the optimal single-photon detunings $\delta_{12}$ and $\delta_{23}$, taking into account realistic ion trajectories and laser phase noise.

Since under those conditions the OBE cannot be solved in a closed form, we integrate them numerically between $t=0$ and $t=t_{\rm max}$. We use a 4$^{th}$ order Runge-Kutta method with a short enough time step (10$^{-9}$ to 5.~10$^{-8}$~s) to well represent the relevant characteristic frequencies of the problem. The initial conditions are $\rho_{11}$=1, and zero for all the other density matrix elements.  Since we consider an open three-level system, the stationary solution is not relevant. The populations and coherences only have a transient behavior and vanish for long times. The signal, i.e. the dissociated fraction is given by $\rho_{44}^\infty=\rho_{44}(t\rightarrow\infty)$; the integration time $t_{\rm max}$ has to be chosen long enough to get a precise estimate of $\rho_{44}^\infty$.

\section{Results}\label{sec_results}

The REMPD signal given by the dissociated fraction $\rho_{44}$ is first studied in Sec.~\ref{subsect_oscill_ion} in the simple case of noiseless lasers and of a single molecular ion with a harmonic motion to characterize sideband effects and identify the Lamb-Dicke regime. In Sec.~\ref{subsect_real_motion}, we come to a more realistic model by including actual ion trajectories to simulate the experimental signal and determine optimal conditions for REMPD signal observation. The OBE results are compared to a simple rate equation model introduced in Sec.~\ref{sec_rate_eq_model}. Finally, we evaluate light shifts and power broadening, and analyze the effects of laser phase noise in Sec.~\ref{sec_systematic_effects}.

\subsection{Single-frequency oscillating ion}\label{subsect_oscill_ion}

\begin{figure}%[!h]
\includegraphics[width=8cm]{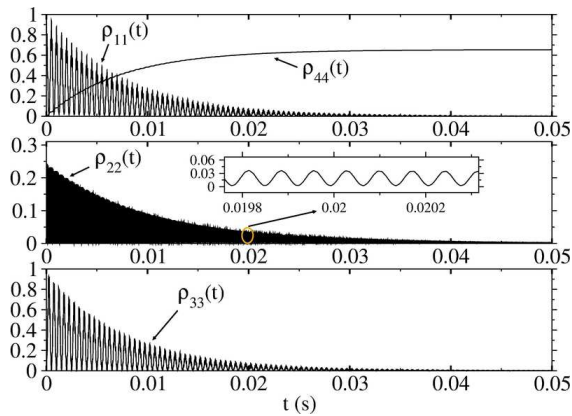}\\
\caption{Time evolution of the populations in the case of a single ion undergoing pure harmonic motion along the z axis. $\Omega_{\rm vibr} = 2 \pi\times 600$~kHz, velocity amplitude: $\tilde{v}=1$~m/s, $\Omega_{12}=\Omega_{23}=2\pi\times 5$~kHz, small detuning $\delta_{12}=-\delta_{23}=2\pi\times 10$~kHz, integration time step: 10$^{-9}$~s.}\label{fig2_time_evolution_small_delta}
\end{figure}

\begin{figure}%[!h]
\includegraphics[width=8cm]{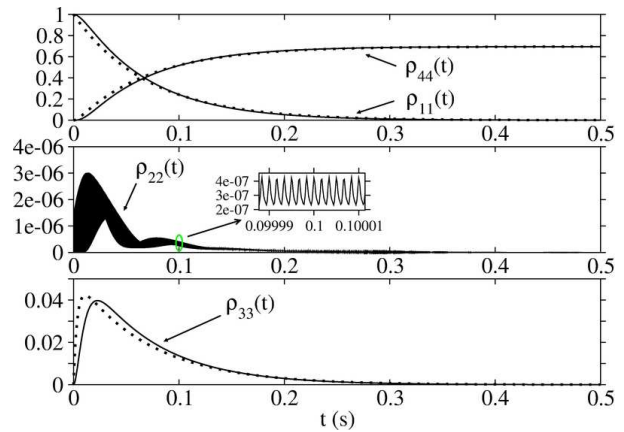}
\caption{Same as Fig.~\ref{fig2_time_evolution_small_delta} but with a large detuning $\delta_{12}=-\delta_{23}=2\pi\times 5$~MHz. Dotted lines are obtained from Eq.~(\ref{eq_rate_model_result_t}) without adjustable parameter.}\label{fig3_time_evolution_large_delta}
\end{figure}

Here, we consider a single ion oscillating with an angular frequency $\Omega_{\rm vibr}$ and velocity amplitude $\tilde{v}$. Figure~\ref{fig2_time_evolution_small_delta} (resp.~\ref{fig3_time_evolution_large_delta}) shows the typical time evolution of the populations $\rho_{11}$, $\rho_{22}$, $\rho_{33}$ as well as the dissociated fraction $\rho_{44}$ in the case of an ion with a pure oscillatory motion for opposite small (large) detunings of 10~kHz (5~MHz) as compared to the single-photon Doppler width $\tilde{v}/\lambda$=~714~kHz. The other parameters of the calculation (see figure captions) correspond to the typical values used throughout the paper. Although the final dissociated fractions $\rho_{44}$ are comparable, the two figures corresponds to completely different conditions.

For small detunings, two-photon excitation is a sequential process involving a large intermediate state population $\rho_{22}$. $\rho_{11}$ and $\rho_{33}$ (resp. $\rho_{22}$) exhibit strong oscillations at 2~kHz (resp. 12 and 14~kHz), see Fig.~\ref{fig2_time_evolution_small_delta}. We have checked that those evolution frequencies are consistent with the generalized Rabi frequencies that can be determined by solving the OBE analytically for an ion at rest ($\bf{\dot{r}}(t)=0$)).

In the large detuning regime (Fig.~\ref{fig3_time_evolution_large_delta}), $\rho_{22}$ always remains negligible, and level $|3\rangle$ is directly excited from level $|1\rangle$ by a two-photon process. Comparing the time scales in Figs.~\ref{fig2_time_evolution_small_delta} and \ref{fig3_time_evolution_large_delta}, one can see that the two-photon process is much slower than the low-detuning sequential process; nonetheless it also leads to a large dissociated fraction after a long enough time. The behavior of $\rho_{11}$ and $\rho_{44}$ in Fig.~\ref{fig3_time_evolution_large_delta} is very close to exponential decay, which will allow to describe the evolution by an effective REMPD rate. The apparent thickness of the $\rho_{22}$ curve is due to fast modulation at the ion oscillation frequency.

\begin{figure}%[!h]
\includegraphics[width=8cm]{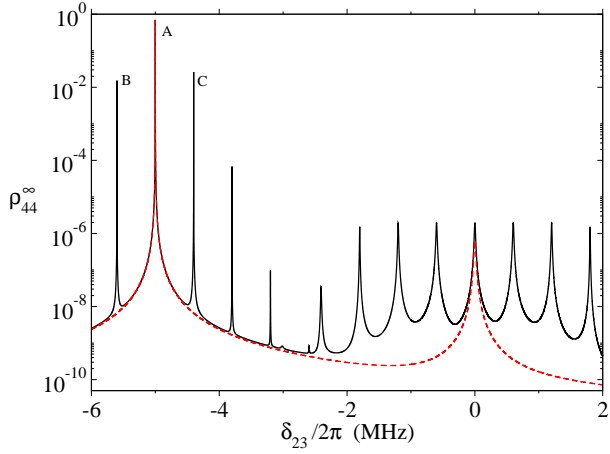}\\
  \caption{Photodissociated fraction as a function of $\delta_{23}$. Red dashed line: ion at rest. Black solid  line: oscillating ion with $\Omega_{\rm vibr}/2\pi=600$~kHz, $\tilde{v}=1$~m/s. Parameters: $\Omega_{12}=\Omega_{23} = 2\pi\times 5$~kHz,  $\delta_{12}/2\pi=5$~MHz. Time step: 5 $\times10^{-8}$~s, $t_{\rm max}=0.5$~s.}
  \label{fig4_spectrum_one_freq}
\end{figure}

We now analyze the REMPD signal $\rho_{44}^\infty$ as a function of $\delta_{23}$ for a (fixed) large detuning $\delta_{12}$. Figure~\ref{fig4_spectrum_one_freq} shows the spectrum for an oscillating ion with $\Omega_{\rm vibr}=2\pi\times 600$~kHz and $\tilde{v}=1$~m/s (the red dashed line is obtained for an ion at rest for comparison). It exhibits two groups of peaks having a sideband structure, in which the sidebands are generated by the Doppler effect due to the ion oscillation, leading to a comb of lines separated by $\Omega_{\rm vibr}$. They correspond to two different processes.

The right part of Fig.~\ref{fig4_spectrum_one_freq}, centered at $\delta_{23}=0$, corresponds to sequential excitation. Since the detuning $\delta_{12}$ is large as compared to the single-photon Doppler width, sequential excitation is inefficient leading to very small dissociated fractions of the order of 10$^{-6}$. In its rest frame, the oscillating ion sees phase modulated laser spectra with a modulation index of $2\pi\tilde{v}/(\lambda \Omega_{\rm vibr})= 1.16$ leading to three significant sidebands on each side of the carrier explaining the broad signal sideband structure.

\begin{figure}%[!h]
\includegraphics[width=8cm]{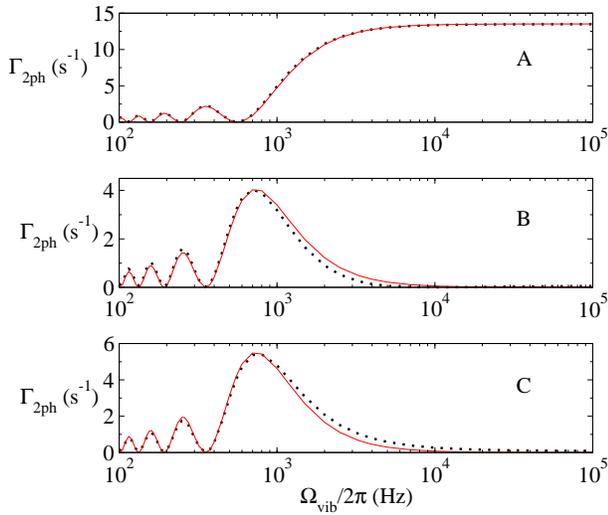}\\
  \caption{Black dotted curves: Two-photon transition rates on the carrier (A) and first sidebands (B,C) of the two-photon resonance versus the oscillation frequency, obtained by solving the OBE for an harmonic oscillation. Red solid curves: same, but obtained using Eqs.~(\ref{eq_proba_2ph_0}) and (\ref{eq_matrix_elem_LambDicke}) of Appendix~\ref{ap_2ph_proba}. Parameters: $\delta_{12}/2\pi=5$~MHz, $\Omega_{12}=\Omega_{23} = 2\pi\times 5$~kHz, $\tilde{v}=1$~m/s. Time step: 10$^{-9}$s, $t_{\rm max}=1$~s. }
  \label{fig5_lamb-dicke-evidence}
\end{figure}

The left part of Fig.~\ref{fig4_spectrum_one_freq}, centered at the two-photon resonance $\delta_{23}=-\delta_{12}$, is the signal due to instantaneous two-photon excitation. It exhibits an intense narrow peak as well as sidebands. However, the sidebands are much smaller than the carrier and drop off very rapidly with sideband order, evidencing the Lamb-Dicke regime. In order to get a more quantitative understanding, we varied the ion oscillation frequency $\Omega_{\rm vibr}$ for a given velocity amplitude ($\tilde{v}=1$~m/s) and determined the two-photon transition rate $\Gamma_{\rm 2ph}$ by fitting the decay of $\rho_{11}(t)$ with Eq.~(\ref{eq_rate_model_result_t}) (see Sec.~\ref{sec_rate_eq_model}), for the carrier and first sidebands of the two-photon signal (peaks A, B and C in Fig.~\ref{fig4_spectrum_one_freq}). Figure~\ref{fig5_lamb-dicke-evidence} shows $\Gamma_{\rm 2ph}$ versus $\Omega_{\rm vibr}$. Red solid lines are obtained from the model given in Appendix~\ref{ap_2ph_proba} (Eqs.~(\ref{eq_proba_2ph_0}-\ref{eq_matrix_elem_LambDicke})). In Eq.~(\ref{eq_matrix_elem_LambDicke}), the effective quantum number $n$ depends on $\Omega_{\rm vibr}$ through the relationship $(n+1/2)\hbar\Omega_{\rm vibr}\approx m\tilde{v}^2/2$, and we used $s=0,\pm 1$ for the carrier A and sidebands B, C respectively. Both approaches are in good agreement, and demonstrate that the system is deep in the Lamb-Dicke regime.

To conclude on the spectrum of Fig.~\ref{fig4_spectrum_one_freq}, let us stress again the important differences with respect to the gas case. In a dilute gas, the velocity can be considered as constant during the interaction with light; as a result, sequential transitions are Doppler-free because the first transition selects a velocity class~\cite{Bjorkholm1976}. This effect does not take place in ion traps, where the ion velocities oscillate with time, and sequential transitions are Doppler-broadened. On the contrary, instantaneous transitions which are Doppler-free in ion traps due to the Lamb-Dicke effect, exhibit residual Doppler broadening in a gas.

\begin{figure}%[!h]
\includegraphics[width=8cm]{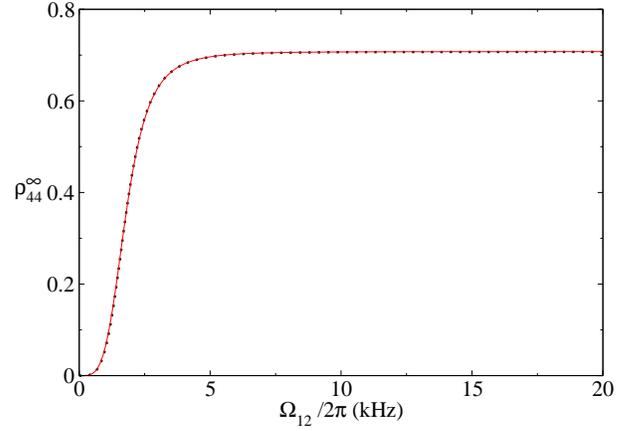}
\caption{Photodissociated fraction versus Rabi frequencies for a detuning $\delta_{12}=-\delta_{23}=2\pi\times5$~MHz. $\Omega_{12}$ and $\Omega_{23}$ are taken equal. The red solid line corresponds to the prediction of Eq.~(\ref{eq_rho44_infty}), and the dotted curve is obtained by solving the OBE with a time step of 10$^{-8}$~s, and $t_{\rm max}=10$~s.}
\label{fig6_rho44_vs_Rabi}
\end{figure}

Figure~\ref{fig6_rho44_vs_Rabi} shows the signal at two-photon resonance as a function of the Rabi frequencies $\Omega_{12}$ and $\Omega_{23}$, assuming that they are equal. The saturation intensity (for which the signal is equal to half its maximum value) is found to correspond to Rabi frequencies of about 2~kHz, in excellent agreement with the rate equation model of Sec.~\ref{sec_rate_eq_model} (Eq.~(\ref{eq_rabi_saturation})). For most of the calculations hereafter, the Rabi frequencies are set to 5~kHz to achieve large signals.

\subsection{Real ion motion}\label{subsect_real_motion}

\begin{figure}%[!h]
\includegraphics[width=8cm]{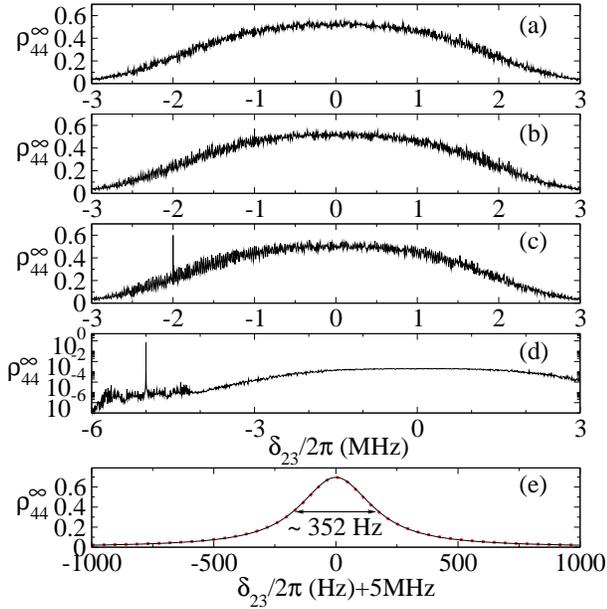}
\caption{
(a) to (d) Dissociated fraction as a function of $\delta_{23}$ for four values of $\delta_{12}/2\pi$: 0, 1, 2 and 5 MHz. (e) Zoom on the Doppler-free peak for $\delta_{12}/2\pi=5$~MHz; the solid line is obtained from Eq.~(\ref{eq_rho44_infty}). The simulations are performed for 20~HD$^+$ ions in a 400~Be$^+$ ion cloud. Parameters: $\Omega_{12} = \Omega_{23}=2\pi\times 5$~kHz, time step 10$^{-8}$~s, $t_{\rm max}=0.5$~s for (a-d) and 10~s for (e).}
\label{fig7_signal_detuning}
\end{figure}

In this Section, we come to a more realistic description of the REMPD dynamics by inserting into the OBE ion trajectories obtained by numerically simulating the motion of 20~HD$^+$ ions sympathetically cooled by 400~Be$^+$ ions (Appendix~\ref{ap_ion_dynamics}). The dissociated fraction is computed for each trajectory and the results are averaged. Figure~\ref{fig7_signal_detuning} shows the dissociated fraction $\rho_{44}$ as a function of detuning $\delta_{23}$ for $\delta_{12}=0$, 1, 2 and 5 MHz. For small detunings, $\rho_{44}$ is dominated by the sequential contribution,leading to a wide Doppler-broadened spectrum which obscures the  Doppler-free instantaneous two-photon signal. For detunings larger than the single-photon Doppler width, the sequential contribution strongly decreases and the narrow Doppler-free peak dominates.

\begin{figure}%[!h]
\includegraphics[width=8cm]{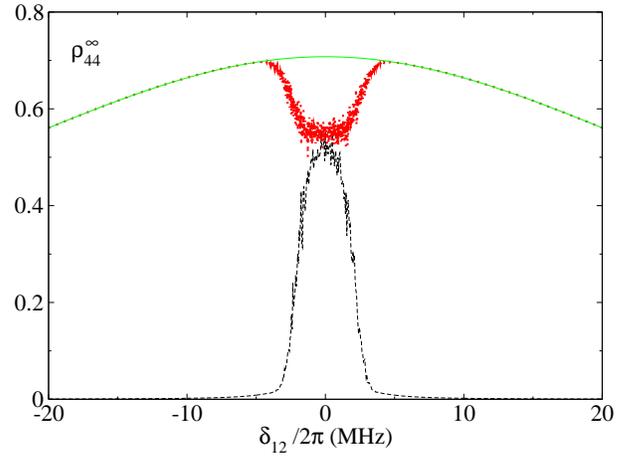}
\caption{Red: Photodissociated fraction $\rho_{44}$ as a function of the detuning $\delta_{12}$ at two-photon resonance ($\delta_{23}=-\delta_{12}$) obtained by solving the OBE. Black: Photodissociated fraction due to sequential two-photon excitation ('noise floor'), which we evaluate at the pedestal of the Doppler-free peak. The green curve is obtained from Eq.~(\ref{eq_rho44_infty}). Parameters: $\Omega_{12}=\Omega_{23}=2\pi\times 5$~kHz, time step 10$^{-8}$~s, $t_{\rm max}=2$ s.}
\label{fig8_signal_to_noise}
\end{figure}

The sequential contribution thus appears as a noise floor that limits the visibility of the Doppler-free signal. In order to determine how close to the resonance the detuning can be set, we compare the Doppler-free signal to the sequential contribution by plotting in Fig.~\ref{fig8_signal_to_noise} the top of the Doppler-free peak and the estimated 'noise floor' due to the sequential signal. The results show that an optimal visibility of the Doppler-free signal is achieved for detunings around 5~MHz, which corresponds to the maximum single-photon Doppler shift experienced by the ions.

%\vspace{1cm}

\subsection{Rate equation model}\label{sec_rate_eq_model}

The analysis of the signal predicted by solving numerically the OBE showed that the optimum value of the detuning is slightly larger than the Doppler width. In that case, the population of level $|2\rangle$ always remains negligible, and the OBE describing the evolution of the three-level system in interaction with the laser fields can be simplified by introducing the two-photon transition probability $\Gamma_{\rm 2ph}$ between levels $|1\rangle$ and $|3\rangle$ (see Appendix~\ref{ap_2ph_proba}). The time evolution of the populations $\rho_{11}$, $\rho_{33}$ and $\rho_{44}$ can then be described by a simple rate equation model. Introducing $\Gamma_3^{\rm eff}=\Gamma_3+\Gamma_{\rm diss}$, the rate equations are written
\begin{equation}\label{eq_rate_equation}
\left\{
\begin{array}{ccl}
\frac{d\rho_{11}}{dt} &=&- (\Gamma_{\rm 2ph} + \Gamma_{1}) \rho_{11} \\
\frac{d\rho_{33}}{dt} &=&  \Gamma_{\rm 2ph}\rho_{11} - \Gamma_3^{\rm eff}\rho_{33}\\
\frac{d\rho_{44}}{dt} &=&  \Gamma_{\rm diss} \rho_{33},
\end{array}
\right.
\end{equation}
where, in order to simplify the expressions, we have replaced $\rho_{11}-\rho_{33}$ by $\rho_{11}$ in the first two equations. This approximation is justified for large detunings, since $\rho_{33}$ then remains much smaller than $\rho_{11}$. The solution corresponding to $\rho_{11}(0)=1$ and $\rho_{33}(0)=\rho_{44}(0)=0$ reads
\begin{equation}
\left\{
\begin{array}{ccl}
\rho_{11}(t) &=& e^{-(\Gamma_1+\Gamma_{\rm 2ph}) t} \\
\rho_{33}(t) &=& \frac{\Gamma_{\rm 2ph}}{\Gamma_3^{\rm eff} - \Gamma_1 - \Gamma_{\rm 2ph}} (e^{-(\Gamma_1+\Gamma_{\rm 2ph})t} - e^{-\Gamma_3^{\rm eff}t}) \\
\rho_{44}(t) &=& \frac{\Gamma_{\rm diss}\Gamma_{\rm 2ph}}{\Gamma_3^{\rm eff}(\Gamma_1+\Gamma_{\rm 2ph})} - \frac{\Gamma_{\rm diss}\Gamma_{\rm 2ph}}{\Gamma_3^{\rm eff}-\Gamma_1-\Gamma_{\rm 2ph}} \\
&& \times \left ( \frac{e^{-(\Gamma_1+\Gamma_{\rm 2ph})t}}{\Gamma_1+\Gamma_{\rm 2ph}} - \frac{e^{-\Gamma_3^{\rm eff}t}}{\Gamma_3^{\rm eff}} \right ).
\end{array}\label{eq_rate_model_result_t}
\right.
\end{equation}
Dotted lines in Fig.~\ref{fig3_time_evolution_large_delta} are plotted from Eq.~(\ref{eq_rate_model_result_t}). They compare very well with the numerical result obtained with an oscillating ion in the large detuning limit, indicating that the instantaneous two-photon contribution is insensitive to the ion motion as expected in the Lamb-Dicke regime. The long-term behavior of $\rho_{44}$ is given by
\begin{equation}
\rho_{44}^\infty = \frac{\Gamma_{\rm diss}\Gamma_{\rm 2ph}}{(\Gamma_3 + \Gamma_{\rm diss})(\Gamma_1+\Gamma_{\rm 2ph})}.
\end{equation}
If $\Gamma_{\rm 2ph} \gg \Gamma_1$ we have simply $ \rho_{44}^\infty \approx \Gamma_{\rm diss}/(\Gamma_{3}+\Gamma_{\rm diss})$. Indeed, in that case, direct losses from level $|1\rangle$ are negligible as compared to excitation to level $|3\rangle$, and $\rho_{44}^\infty$ is given by the branching ratio between dissociation and natural relaxation.

In the general case, replacing $\Gamma_{\rm 2ph}$ with the expression given by Eq.~(\ref{eq_proba_2ph_simple}), we obtain an expression for the photodissociated fraction that is valid in the Lamb-Dicke regime:
\begin{equation}
\rho_{44}^\infty = \frac{\Gamma_{\rm diss}}{\Gamma_{1}} \frac{\Omega_{12}^2 \Omega_{23}^2}{\delta_{12}^2} \frac{1}{\delta_{13}^2 + \frac{(\Gamma_3^{\rm eff})^2}{4} + \frac{\Gamma_3^{\rm eff}}{\Gamma_1} \frac{\Omega_{12}^2 \Omega_{23}^2}{\delta_{12}^2}  }.\label{eq_rho44_infty}
\end{equation}
Figure~\ref{fig6_rho44_vs_Rabi} showing $\rho_{44}^\infty$ versus the Rabi frequencies is obtained for an oscillating ion in the large detuning limit. Again, the results of Eq.~(\ref{eq_rho44_infty}) closely match the OBE numerical model.

The saturation Rabi frequency, defined as the Rabi frequency product $\Omega_{12}\Omega_{23}$ for which $\rho_{44}^\infty=\Gamma_{\rm diss}/2\Gamma_3^{\rm eff}$, is given by
\begin{equation} \label{eq_rabi_saturation}
\Omega_{12}\Omega_{23}=\delta_{12}\sqrt{\frac{\Gamma_1}{\Gamma_3^{\rm eff}}(\delta_{13}^2+(\Gamma_3^{\rm eff})^2/4)},
\end{equation}
which reduces to $\Omega_{12}\Omega_{23}=\delta_{12}\sqrt{\Gamma_1 \Gamma_3^{\rm eff}}/2$ on two-photon resonance.

%Interpretation is still missing !! \textit{\textbf{LH partially answers JK : It seems that the power broadening goes to zero on double resonance ($\delta_{12}=0$). The analytical model is not valid on double resonance, so $\delta_{12}=0$ should not be set to 0.}}

Comparing the green solid line in Fig.~\ref{fig8_signal_to_noise} with the red line representing the solution of the OBE shows that the rate equation model accurately predicts $\rho_{44}^\infty$ for detunings larger than the Doppler width, but as expected, fails for small detunings. Finally, the Doppler-free line obtained by solving the OBE and shown in Fig.~\ref{fig7_signal_detuning}(e) has a Lorentzian shape of amplitude 0.7 and FWHM 352~Hz  in excellent agreement with the predictions of Eq.~(\ref{eq_rho44_infty}) giving 0.71 for the amplitude and 354~Hz for the width.

\subsection{Systematic shifts and line broadening}\label{sec_systematic_effects}

In this Section, we study the main effects that may perturb the Doppler-free REMPD signal, i.e. lightshifts, power broadening and laser frequency noise. Only the large-detuning limit will be studied, and numerical results obtained from the OBE will be compared with predictions of the simple analytical model developed in Sec.~\ref{sec_rate_eq_model}.

\subsubsection{Light Shifts}

The light shift experienced by the lower and upper levels $|1\rangle$ and $|3\rangle$ are given by $+\Omega_{12}^2/\delta_{12}$ and $-\Omega_{23}^2/\delta_{23}$, respectively~\cite{Liao1975,Salomaa1977}. Close to the two-photon resonance defined by $\delta_{12}=-\delta_{23}$, both shifts have the same sign leading to a compensated light shift for the transition frequency:
\begin{equation}
\Delta_{LS}=(\Omega_{23}^2-\Omega_{12}^2)/\delta_{12}.\label{eq_light_shift}
\end{equation}
As was shown in Sec.~\ref{subsect_real_motion}, the optimal value of the detuning $\delta_{12}$ is of the order of the Doppler width (a few MHz), whereas the Rabi frequencies are of a few kHz. Therefore the light shift typically amounts to a few Hz, i.e. an relative shift of about 10$^{-14}$ on the transition frequency. Moreover, laser intensities can be chosen in order to get equal Rabi frequencies thus canceling the light shifts.

\begin{figure}%[!h]
\includegraphics[width=8cm]{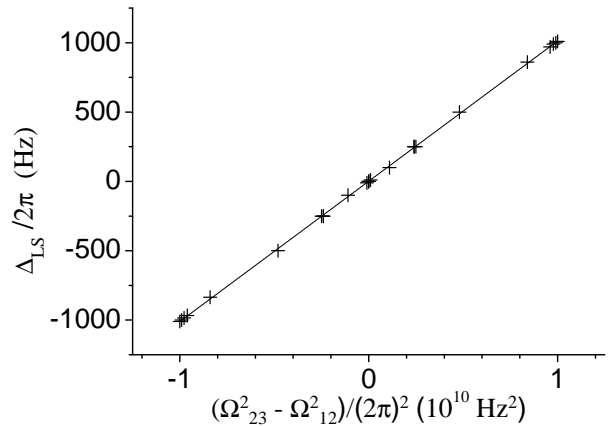}
\caption{Crosses: Light shift of the two-photon resonance versus Rabi frequencies. The two-photon resonance is located by finding the maximum of the Doppler-free peak (see Fig.~\ref{fig7_signal_detuning} (e)). Solid line: linear fit giving a slope of 1.011(2)~10$^{-7}$~Hz/(Hz)$^2$. Parameters: $\delta_{12}=10$~MHz, time step 10$^{-8}$~s, $t_{\rm max}=0.5$~s.}
\label{fig9_light_shifts}
\end{figure}

In Fig.~\ref{fig9_light_shifts}, the position $\delta_{23}$ of the two-photon peak is plotted versus $\Omega_{23}^2-\Omega_{12}^2$ for a fixed detuning $\delta_{12}=10$~MHz. It has a linear dependence with a slope of 1.011(2)~10$^{-7}$~Hz/(Hz)$^2$, in good agreement with Eq.~(\ref{eq_light_shift}) which predicts 10$^{-7}$~Hz/(Hz)$^2$.

\subsubsection{Power Broadening}

\begin{figure}%[!h]
\includegraphics[width=8cm]{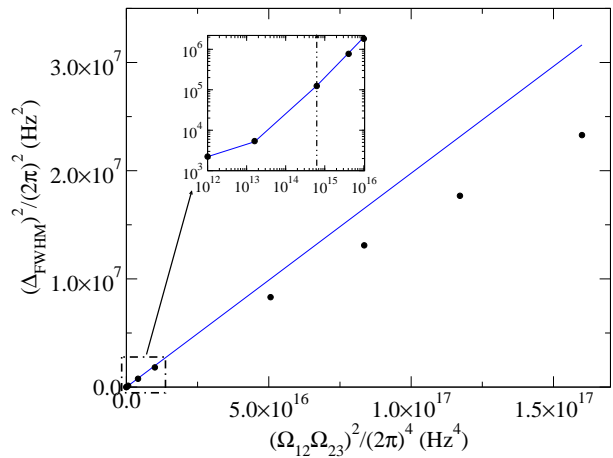}
\caption{Squared  width (FWHM) of the Doppler-free peak versus Rabi frequencies. The solid line is obtained from Eq.~(\ref{eq_rho44_infty}). The vertical dashed line in the inset corresponds to the typical Rabi frequencies used throughout the paper, i.e. $\Omega_{12}=\Omega_{23}=2\pi\times 5$~kHz. Parameters: $\delta_{12}=$~5~MHz, time step 10$^{-8}$~s, $t_{\rm max}=20$~s.}
\label{fig10_rho44_width}
\end{figure}

A simple expression of the power broadening is easily deduced from Eq.~(\ref{eq_rho44_infty}). Figure~\ref{fig10_rho44_width} compares the broadening predicted by Eq.~(\ref{eq_rho44_infty}) to a more precise calculation from the numerical solution of the OBE. The inset shows that there is excellent agreement at low intensity. For very large Rabi frequencies, the numerically obtained power broadening is smaller than expected from Eq.~(\ref{eq_rho44_infty}). This discrepancy stems from the fact that Eq.~(\ref{eq_rho44_infty}) is obtained using Eq.~(\ref{eq_proba_2ph_simple}) for $\Gamma_{\rm 2ph}$, which is valid if $\Gamma_{\rm 2ph}\ll\Gamma_3^{\rm eff}$ but not for large laser fields.

As already mentioned in Sec.~\ref{subsect_real_motion}, the FWHM of the two-photon peak for the laser intensities used throughout the paper ($\Omega_{12}=\Omega_{23}=2\pi\times 5$~kHz, signalled  by a vertical dashed line in the inset of Fig.~\ref{fig10_rho44_width}), is 354~Hz (see Fig.~\ref{fig7_signal_detuning}(e)) while the prediction of Eq.~(\ref{eq_rho44_infty}) is 351~Hz. Thus power broadening significantly degrades the resolution with respect to the effective linewidth $\Gamma_3^{\rm eff} = 45$~Hz. Lower intensities can be used to improve the resolution, at the cost of a slightly reduced signal-to-noise ratio.

\subsubsection{Laser frequency noise}

The analysis of light shifts and power broadening shows that REMPD spectroscopy at the sub-kHz level is feasible. In this Section, we discuss the effect of the laser width on the signal, using  both numerical solutions of OBE and an analytical model.

So far, we have assumed noiseless laser fields by setting $\varphi(t)=\varphi'(t)=0$ in Eq.~(\ref{eq_E_field}). This means that the two laser fields are supposed to be perfectly phase-locked.
The discussion in Sec.~\ref{sec_time_freq_scale} shows that laser linewidths cannot always be neglected as compared to Rabi frequencies and level widths. The REMPD experiment involves frequency controlled diode laser sources, which have a Lorentzian lineshape. In order to include the laser frequency noise into the model, two independent noisy phases $\varphi(t)$ and $\varphi'(t)$ are numerically generated as centered gaussian stationary processes with the desired shape and width~\cite{Elliott1982,DiDomenico2010} as explained in Appendix~\ref{ap_laser_phase_noise}, and used as inputs in the OBE~(\ref{eq_OBE}). For both lasers the phase noise bandwidth $B$ is chosen to be 100~kHz and the width $\Delta_{\rm FWHM}$ is varied from a few Hz to 30~kHz.

\begin{figure}%[!h]
\includegraphics[width=8cm]{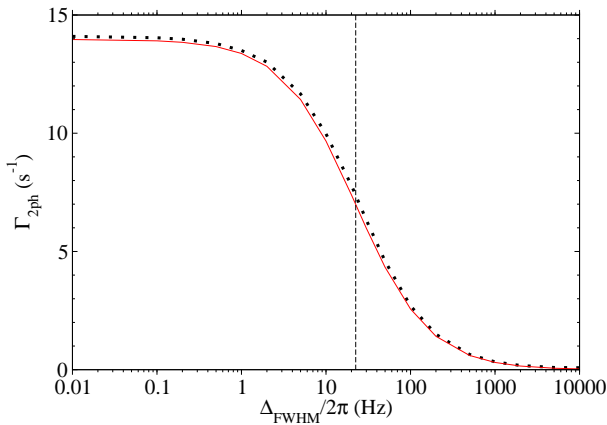}
\caption{Two-photon transition rate versus the laser width (assuming a lorentzian spectrum). The black dotted curve is obtained by integrating the OBE with a time step of 10$^{-8}$ and $t_{\rm max}=0.5$~s. The red solid line is obtained from Eq.~(\ref{eq_NoiseEffect}). The dashed vertical line corresponds to $\Delta_{\rm FWHM}=\Gamma_3^{\rm eff}/2$. Parameters: $\Omega_{12}=\Omega_{23}=2\pi\times 5$~kHz, $\delta_{12}=-\delta_{23}=2\pi\times 5$~MHz.}
\label{fig11_phase_noise_effect}
\end{figure}

The effect of laser phase noise on two-photon transition rates is theoretically addressed in~\cite{Mollow1968}. For two uncorrelated phases $\varphi(t)$ and $\varphi'(t)$, formula~(\ref{eq_proba_2ph}) is modified into
\begin{equation}\label{eq_NoiseEffect}
\Gamma_{\rm 2ph} =   \frac{\Omega_{12}^2 \Omega_{23}^2}{\delta_{12}^2} \frac{\Gamma_3^{\rm eff}+ 2 \Delta_{\rm FWHM}}{\delta_{13}^2 + (\Gamma_3^{\rm eff}+ 2 \Delta_{\rm FWHM})^2/4}
\end{equation}
Just like Eq.~(\ref{eq_proba_2ph}), the above expression is valid in the large-detuning limit $\delta_{12}> \Gamma_D$. In Fig.~\ref{fig11_phase_noise_effect}, we plot $\Gamma_{\rm 2ph}$ versus $\Delta_{\rm FWHM}$ assuming both lasers have the same width. Numerical results from the OBE are in very good agreement with Eq.~(\ref{eq_NoiseEffect}), and show that it is desirable to have laser widths smaller than the effective width $\Gamma_3+\Gamma_{\rm diss}$ of the upper level in order not to limit the two-photon transition rate, as well as the resolution.

%\vspace{1cm}

\subsubsection{Effects of micromotion}\label{sec_micromotion}
In a Paul trap, the ions undergo micromotion driven by the RF field at the ion location $(x,y,z)$.
In this section, we evaluate the magnitude of the micromotion in the linear trap described in~\cite{Koelemeij2012} to show
that it has a negligible impact on the two-photon lineshape and that the associated second order Doppler effect
does not limit the expected resolution.

The linear trap geometry is defined by an effective inner radius $r_0$=3.5~mm and is operated using a RF voltage $V_0$=270~V at $\Omega_{RF}$=2$\pi\times$13.3~MHz resulting in a $\omega_r$=2$\pi\times$0.9~MHz HD$^+$ radial trap frequency. The micromotion amplitude $\delta{\bf r}$ is linked to the RF field ${\bf E}_{RF}$ by
$\delta{\bf r}=-q{\bf E}_{RF}/(m\Omega_{RF}^2)$. The leading  components of the RF field are ${\bf E}_{RF}=(-V_{0}\ x/r_0^2,V_{0}\ y/r_0^2,E_{RF,z})\cos(\Omega_{RF}t)$. The radial components correspond to the trap's quadrupolar field. The axial component is a worst case value, obtained using a finite difference analysis (SIMION software) to model the actual trap potential taking into account the maximum possible deviation of end cap electrodes from the ideal geometry; $E_{RF,z}$ is less than 100~V/m over the ion cloud extension.
However, trap imperfections, RF phase differences on the trap electrodes and stray electric fields may lead to excess micromotion, which in turn can give rise to second-order Doppler shifts of the observed transition frequency, as well as additional sideband features in the spectrum~\cite{Berkeland1998,Pyka2013}. Stray electric fields may be compensated by applying voltages on the trap electrodes to position the Be$^+$ and HD$^+$ ion clouds symmetrically with respect to the trap axis. From the applied voltages and the trap geometry, the residual stray field amplitude is estimated to be smaller then 7.3~V/m. The maximum radial displacement $r_{max}^{rad}$ is obtained by balancing the stray electric force $q E_{stray}$ with the ponderomotive force $m\omega_r^2r_{max}^{rad}$, leading to
$r_{max}^{rad}=q E_{stray}/(m\omega_r^2)$=7.3~$\mu$m and maximum radial RF field components of 114~V/m. The maximum axial and radial micromotion amplitudes $\delta x$, $\delta y$ and $\delta z$ are all less than 0.5~$\mu$m, much smaller than the effective transition wavelength.
Furthermore, the ion trap was designed such that RF phase differences do not exceed 3~mrad. For the above trap parameters, this implies a maximum micromotion amplitude due to RF phase differences of $0.4~\mu$m~\cite{Berkeland1998}.

Micromotion might lead to sidebands in the two-photon excitation spectrum, located $\pm$13.3~MHz from the main spectral feature. Nevertheless, under the present conditions, the modulation index $|({\bf k}-{\bf k}')\cdot\delta{\bf r}|<$0.007 is very small leading to strongly suppressed sidebands, justifying ignoring micromotion in the interaction model.

Although the micromotion amplitude is small, the associated velocity amplitude is large and second order Doppler shift and broadening have to be evaluated. It is given by $\delta f/f=-\left<v(t)^2\right>/(2c^2)$. For micromotion with amplitude $\delta{\bf r}$=0.9~$\mu$m, it is given by $-(\delta{\bf r})^2\Omega_{RF}^2/(4c^2)=$-1.5$\times$10$^{-14}$. Including RF phase differences, the shift may reach -1.8$\times$10$^{-14}$ corresponding to less than 4~Hz on individual laser frequencies. This is much smaller than the expected two-photon linewidth and cannot hinder the two-photon line observation. Nevertheless, careful micromotion compensation is necessary to reach the 10$^{-14}$ accuracy level.

\section{Influence of BBR on REMPD and signal strength}\label{sec_BBR}

In the preceding sections, the photodissociated fraction was interpreted as the spectroscopic signal of interest. However, in previous experiments spectroscopic signals were obtained by comparing the initial number of trapped HD$^+$ ions, $N_i$, to the remaining number of HD$^+$ ions after REMPD, $N_f$, by constructing a signal $s=(N_i - N_f)/N_i$~\cite{Koelemeij2007,Koelemeij2012}. Obviously, the finite size of the trapped HD$^+$ samples may lead to additional saturation effects. It should also be noted that before REMPD, most of the HD$^+$ ions are in states other than $v=0, L=3$ as the ambient BBR (temperature $T=300$~K) distributes population over rotational states with $L=0$ to $L=6$~\cite{Koelemeij2007b}. Each rotational level is furthermore split into four ($L=0$), ten ($L=1$), or twelve ($L\ge 2$) hyperfine states. As a consequence, only a few per cent of the HD$^+$ ions may be found to be in a particular hyperfine state. For example, 2.6~\% of the HD$^+$ ions are in the favored initial hyperfine state with $(v,L)=(0,3)$ and $(F,S,J)=(1,2,5)$ (see Appendix~\ref{ap_Zeeman effect}). At first glance, one would therefore not expect to achieve a signal $s$ larger than 0.026, which is barely above the noise background observed by Koelemeij \textit{et al.}~\cite{Koelemeij2007}. However, for REMPD durations on the order of 1~s or longer, redistribution of population by BBR becomes an important factor, as this takes place on a similar timescale. In fact, BBR will continue to refill the initial state population while it is being depleted via REMPD, thereby enhancing the signal $s$. To estimate the expected signal strength, we treat the interaction of the ensemble of HD$^+$ ions with BBR and the REMPD lasers in the form of Einstein rate equations, which we integrate over the REMPD duration, $t$, to obtain $s(t)$. Here we introduce two simplifying assumptions: first, the REMPD process is considered sufficiently efficient so that no spontaneous emission from high vibrational states occurs. Second, all HD$^+$ ions are considered to be in states with $v=0$ and $L=0 ... 5$ (we ignore the population in $L=6$, which is less than 2~\%. Taking hyperfine structure into account, the rate equations read
\begin{eqnarray}\label{eq:EinsteinRE}
\frac{\text{d}}{\text{d}t} &&\rho_{\alpha L}
= \sum_{\alpha'} \left( A^{\alpha' L+1}_{\alpha L} + B^{\alpha' L+1}_{\alpha L} W(\omega^{\alpha' L+1}_{\alpha L}, T)\right)\rho_{\alpha' L+1} \nonumber \\
& & +\sum_{\alpha'}  B^{\alpha' L-1}_{\alpha L} W(\omega^{\alpha L}_{\alpha' L-1},T)  \rho_{\alpha' L-1} \\
& & -\sum_{\alpha'} \left( A^{\alpha L}_{\alpha' L-1} + B^{\alpha L}_{\alpha' L-1} W(\omega^{\alpha L}_{\alpha' L-1},T)\right)\rho_{\alpha L} \nonumber \\
& & -\sum_{\alpha'} B^{\alpha L}_{\alpha' L+1} W(\omega^{\alpha' L+1}_{\alpha L},T)\rho_{\alpha L} - \delta_{\alpha \alpha_0} \delta_{L L_0} \Gamma_{\rm 2ph}\rho_{\alpha L}.\nonumber
\end{eqnarray}
Here, the hyperfine populations $\rho$ are labeled by the hyperfine index $\alpha \equiv (F,S,J)$. Transition frequencies are written as $\omega^{\alpha' L'}_{\alpha L}$, where the upper and lower indices refer to the upper and lower levels, respectively. The BBR spectral energy density at temperature $T$ is denoted as $W(\omega, T)$. The hyperfine state subject to REMPD at rate $\Gamma_{\rm 2ph}$ is labeled by $\alpha_0$ and $L_0$. Introducing the equivalent notation $A_{ij}=A^i_j=A^{\alpha L}_{\alpha' L'}$ (and likewise for $B_{ij}$ and $\omega_{ij}$), the rate coefficients for spontaneous emission from an upper state $i$ to a lower state $j$ are written as
\begin{equation}
A_{ij}=\frac{\omega_{ij}^3}{3 \pi \epsilon_0 \hbar c^3} \frac{\mathcal{S}_{ij}}{2 J_i+1} \mu_{ij}^2.
\end{equation}
The radial dipole matrix elements $\mu_{ij}$ are those presented previously in Ref.~\cite{Koelemeij2011}, and the hyperfine line strengths $\mathcal{S}_{ij}$ are derived in a similar fashion as in Refs. \cite{Spezeski1977,Bakalov2011}. The calculation of $\mathcal{S}_{ij}$ involves hyperfine eigenvectors, which are obtained by diagonalization of an effective spin Hamiltonian~\cite{Bakalov2006}. Spin coefficients for $v=0, L=0...4$ are taken from \cite{Bakalov2006}, and extrapolation of these coefficients results in a set of spin coefficients for $v=0, L=5$. Likewise, the rate coefficients for stimulated emission and stimulated absorption are
\begin{equation}
B_{ij}=\frac{\pi^2 c^3}{\hbar \omega_{ij}^3} A_{ij}
\end{equation}
and
\begin{equation}
B_{ji}=\frac{2J_i+1}{2 J_j+1} B_{ij},
\end{equation}
respectively. After integrating Eq.~(\ref{eq:EinsteinRE}) to obtain $\rho_{\alpha L}(t)$ as a function of the REMPD duration $t$, the signal $s(t)$ becomes
\begin{equation}
s(t)=\frac{\sum_{\alpha,L}\rho_{\alpha L}(0)-\rho_{\alpha L}(t)}{\sum_{\alpha,L}\rho_{\alpha L}(0)}.
\end{equation}
Here, the initial distribution of populations $\rho_{\alpha L}(0)$ is assumed to be a thermal distribution corresponding to the temperature of the BBR (which is assumed to be 300~K here).

\begin{figure}[!ht]
  \centering
  \includegraphics[scale=0.55]{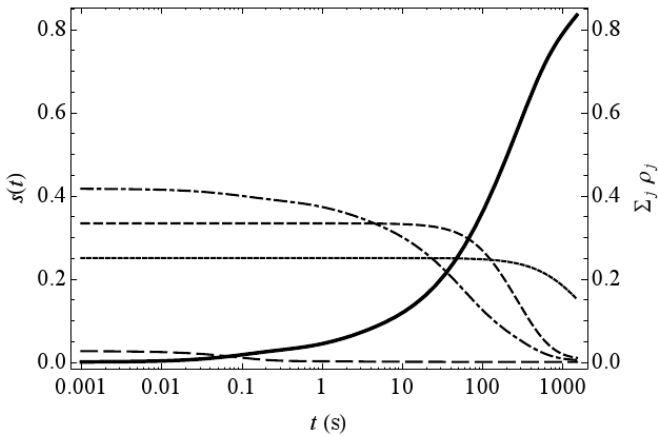}
  \caption{Log-linear plot of the signal strength $s(t)$ as a function of REMPD duration $t$, for two-photon transitions starting from the hyperfine state with $(L,F,S,J)=(3,1,2,5)$ (solid curve) in $v=0$. Shown as well are populations of certain 'spin classes' (rightmost vertical axis). Long-dashed curve, population in $(L,F,S,J)=(3,1,2,5)$; dash-dotted curve, sum over $L,J$ of all population in states with $F=1, S=2$; short-dashed curve, sum over $L,S,J$ of all population in states with $F=1, S \neq 2$; dotted curve, sum over $L,S,J$ of all population in states with $F=0$.}
  \label{fig:SigsBBR}
\end{figure}

We compute signal strengths for the conditions of Fig.~\ref{fig11_phase_noise_effect}, and for a laser linewidth of 10~Hz, for which the REMPD rate is about $10~$s$^{-1}$. The result for the transition starting from the hyperfine level with $(F,S,J)=(1,2,5)$ is shown in Fig.~\ref{fig:SigsBBR}. Different time scales can be identified in the growth of $s(t)$. After 0.2~s, nearly all the population in the initial state $(F,S,J)=(1,2,5)$ is dissociated, and the signal corresponds to the initial hyperfine state population of 0.026. After $t=0.2$~s, BBR continues to replenish population from states with $L \neq 3$ (and with primarily $(F,S)=(1,2)$) through allowed electric-dipole transitions. Transitions between states with equal $F$ but different $S$ are only allowed by virtue of hyperfine mixing and therefore are considerably weaker; transitions between states with different $F$ are even less allowed. The former become important after $t=100$~s, when most HD$^+$ ions with $(F,S)=(1,2)$ have been dissociated, whereas the latter start to dominate the dissociation dynamics only after $t=700$~s when most HD$^+$ ions with $F=1$ have been depleted. The population dynamics are illustrated by the curves in Fig.~\ref{fig:SigsBBR}.

\begin{figure}[!ht]
  \centering
  \includegraphics[scale=0.6]{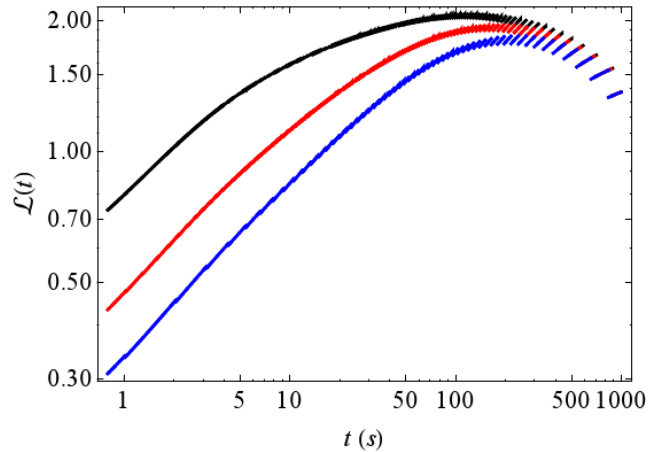}
  \caption{(Color online) Figure-of-merit function $\mathcal{L}$ as a function of REMPD duration $t$, and for various values of the overhead $t_{\rm oh}$. $T_{\rm exp}=3600$~s. Black curve, $t_{\rm oh}=10$~s; red curve, $t_{\rm oh}=30$~s; blue curve, $t_{\rm oh}=60$~s. In all cases, the optimum REMPD duration is near 100~s.}
  \label{fig:optimumREMPD}
\end{figure}

For efficient data acquisition, it is important to find the optimum REMPD duration. Fig.~\ref{fig:SigsBBR} shows that longer durations lead to larger signals. On the other hand, shorter durations allow more data points to be acquired within a given amount of time, $T_{\rm exp}$, which can be averaged to improve the signal-to-noise ratio. The optimum duration depends also on the overhead per data point (\textit{e.g.} time needed to expunge the remaining HD$^+$ ions from the trap, and reload a fresh sample of HD$^+$ ions for the next REMPD cycle). We define a figure of merit for the signal quality, $\mathcal{L}$, obtainable given a total time $T_{\rm exp}$, REMPD duration $t$ and the overhead, $t_{\rm oh}$, as follows. The number of experiments that can be done is $N_{\rm exp} = \lfloor T_{\rm exp}/(t+t_{\rm oh}) \rfloor$, where $\lfloor\, \rfloor$ denotes the floor. Assuming the signal-to-noise ratio improves as $\sqrt{N_{\rm exp}}$, our figure of merit becomes
\begin{equation}
\mathcal{L}(t)=s(t)\sqrt{N_{\rm exp}}=s(t) \sqrt{\lfloor T_{\rm exp}/(t+t_{\rm oh}) \rfloor}.
\end{equation}
$\mathcal{L}(t)$ is plotted for $T_{\rm exp}=3600~s$ and for various values of $t_{\rm oh}$ in Fig.~\ref{fig:optimumREMPD}. Typically, $t_{\rm oh}$ is 30--60~s, for which we find an optimum REMPD duration of~$\sim 100$~s. In this case, we find from Fig.~\ref{fig:SigsBBR} that about 35~\% of the HD$^+$ ions are dissociated. We point out that this is much larger than the 1--2~\% measurement noise observed by Koelemeij \textit{et al.}~\cite{Koelemeij2007}. A spectral lineshape consisting of at least twenty data points may therefore be obtained with a good signal-to-noise ratio within the course of one hour.

\section{Conclusion}

We have shown that Doppler-free signals can be observed on trapped HD$^+$ ions by nearly-degenerate two-photon spectroscopy, taking advantage of a quasi-harmonic three-level ladder in the rovibrational spectrum. The suppression of the Doppler effect, due to an effective Lamb-Dicke regime with respect to the simultaneous absorption of counterpropagating photons, opens the way to high-resolution spectroscopy at the natural width limit. Numerical simulations of the REMPD signal, taking into account saturation effects, realistic ion trajectories and laser phase noise, allowed us to determine the optimal laser detunings, which are slightly larger than the single-photon Doppler width. In this large-detuning limit, the population of the intermediate state may be neglected, and a simplified model of the two-photon transition rate was shown to be in excellent agreement with our numerical results. Finally, blackbody radiation (BBR) redistribution among rovibrational and hyperfine levels was taken into account to get realistic estimates of experimental signal strengths.

With the parameters used in the paper, the predicted linewidth of 350~Hz is dominated by power broadening. It may be reduced to about 100~Hz by using lower intensities, at the cost of a slight diminution of the signal-to-noise ratio. The line center may eventually be determined with about 5~Hz accuracy, corresponding to a relative accuracy of $1\times 10^{-14}$.
Other systematic effects such as quadrupolar shifts, light shifts by cooling and dissociation lasers, BBR shifts, Stark shifts due to stray electric fields and AC Zeeman shifts are estimated to be below 10$^{-15}$, as discussed in a recent study~\cite{Bakalov2013}.

% A very recent study has
%pointed out the presence of level shifts due to the electric quadrupole
%moment of rovibrational states in HD+ due to the quadrupolar electric RF field~\cite{Bakalov2013}. These level shifts are
%estimated to contribute at the level of $1\times 10^{-15}$ only and are not discussed further here.

Potential applications of the proposed spectroscopic method
include improved tests of QED~\cite{Korobov2009,Korobov2013}, an improved determination of the proton-to-electron mass ratio~\cite{Wing1976,Roth2008}, as well as studies of its time variation~\cite{Schiller2005} and searches for possible fifth forces~\cite{Salumbides2013}.

For the rovibrational levels of HD$^+$ selected in this study, the mismatch of the intermediate state is only 0.2\% of the one-photon frequency, leading to a long effective wavelength $\lambda_{\rm eff} = 0.7$~mm. It is worth noting that the effective Lamb-Dicke regime could still be reached with significantly higher frequency mismatch, possibly up to 10\% for excitation wavelengths in the micron range ($\lambda_{\rm eff} \sim 10$~$\mu$m). This means that the proposed method has potential for application to many other molecular (or even atomic) ions, since the existence of such quasi-coincidences is quite probable in a rich rovibrational spectrum characterized by a quasi-harmonic vibrational ladder. In the case of HD$^+$, two other promising transitions are worth pointing out: $(v\!=\!0,L\!=\!3$ to $v\!=\!12,L\!=\!3$ via $v\!=\!5,L\!=\!2$ with wavelengths near 1.18~$\mu$m, and $v\!=\!0,L\!=\!4$ to $v\!=\!16,L\!=\!4$ via $v\!=\!6,L\!=\!3$ near 1.01~$\mu$m~\cite{Moss1993}. Finally, the proposed method could also be extended to multiphoton transitions in a configuration where the laser wavevectors nearly add up to zero~\cite{Biraben1976}.

\appendix

\section{HD$^+$ Zeeman effect}\label{ap_Zeeman effect}

As discussed in Sec.~II, it is preferable to address simultaneously all Zeeman components of the two-photon transition in order to get sufficiently large signals. In the Amsterdam experiment, a static $B$-field is used to define a quantization axis and cool Be$^+$ ions with a single circularly polarized laser beam~\cite{Koelemeij2012}. Experimental investigation showed that the minimal $B$-field value that still enables efficient cooling is about 0.02 G. The Zeeman splitting of the two-photon transition in such a field thus sets a lower limit for the width of the lineshape, which may be broadened as required by control of $\Gamma_\text{diss}$ and the linewidth of the excitation lasers. It is therefore desirable to select a hyperfine component having a low Zeeman effect in order to minimize line broadening and maximize the two-photon transition rate for a given laser intensity. In view of this, It is important to evaluate the Zeeman splitting of the $(v=0,L=3) \rightarrow (v=9,L=3)$ two-photon transition with account of the hyperfine structure, so as to (i) select the most promising hyperfine component, and (ii) determine the optimal dissociation rate and laser linewidth accordingly.

Following the approach of Ref.~\cite{Bakalov2011}, we write the effective spin Hamiltonian for an HD$^+$ ion in a rovibrational state $(v,L)$, with an external magnetic field $\mathbf{B}$ oriented along the $z$ axis:
\begin{eqnarray}
H_{\rm eff}^{tot}&& = H_{\rm eff}^{\rm hfs} + E_{10}(\mathbf{L}\!\cdot\!\mathbf{B}) \\
&&+ E_{11}(\mathbf{S_p}\!\cdot\!\mathbf{B}) + E_{12}(\mathbf{S_d}\!\cdot\!\mathbf{B}) + E_{13}(\mathbf{S_e}\!\cdot\!\mathbf{B})\nonumber, \label{heff}
\end{eqnarray}
where $H_{\rm eff}^{\rm hfs}$ is the effective spin Hamiltonian in the absence of magnetic field derived in~\cite{Bakalov2006}, and
\begin{subequations}
\begin{equation}
E_{10} = -\mu_B \, \sum_i \frac{Z_i m_e}{m_i} \frac{\langle v L || L || v L \rangle}{\sqrt{L(L+1)(2L+1)}}
\end{equation}
\begin{equation}
E_{11} = - \frac{e\mu_p}{m_p c} = -4.257 \, 7 \, \mbox{kHz G}^{-1}
\end{equation}
\begin{equation}
E_{12} = - \frac{e\mu_d}{2 m_d c} = -0.653 \, 9 \, \mbox{kHz G}^{-1}
\end{equation}
\begin{equation}
E_{13} = \frac{e\mu_e}{m_e c} = 2.802 \, 495 \, 3 \, \mbox{MHz G}^{-1}
\end{equation}
\end{subequations}
where 2010 CODATA values of fundamental constants were used. The value of $E_{10}$ is calculated using nonrelativistic variational wavefunctions~\cite{Karr2008b}. We obtain $E_{10} = -0.557 \, 92 \mbox{\ kHz G}^{-1}$ for the $(v=0, L=3)$ level, in agreement with Table 1 of~\cite{Bakalov2011}, and $E_{10} = -0.502 \, 81 \mbox{\ kHz G}^{-1}$ for the $(v=9, L=3)$ level.

In the presence of a magnetic field, the hyperfine states of HD$^+$ labeled with $F$, $S$ and $J$ (see Sec.~II), are split into sub-levels distinguished by the quantum number $M_J$. We diagonalize the Hamiltonian~(\ref{heff}) for $M_J = \pm J$ and $B = 0.02$ G, in order to obtain the Zeeman shifts $\Delta E ^{vLFSJM_J} = E ^{vLFSJM_J}(B) - E ^{vLFSJM_J}(0)$. Results are given in Table~\ref{TableZeeman}.

\onecolumngrid

\begin{table}[h]
\begin{center}
{\scriptsize
\begin{tabular}{cclrrrlrlrrrlrrrrrr}
\hline\hline
 & & \multicolumn{3}{c}{$(F,S) = (0,1)$} && \multicolumn{1}{c}{$(1,0)$} && \multicolumn{3}{c}{$(1,1)$}& & \multicolumn{5}{c}{$(1,2)$} \\
\cline{3-5} \cline{7-7} \cline{9-11} \cline{13-17}  \\
$v$ & $L$ & \multicolumn{1}{c}{$J=4$} & \multicolumn{1}{c}{$J=3$} & \multicolumn{1}{c}{$J=2$} && \multicolumn{1}{c}{$J=3$} && \multicolumn{1}{c}{$J=4$} & \multicolumn{1}{c}{$J=3$} & \multicolumn{1}{c}{$J=2$} & & \multicolumn{1}{c}{$J=5$} & \multicolumn{1}{c}{$J=4$} & \multicolumn{1}{c}{$J=3$} & \multicolumn{1}{c}{$J=2$} & \multicolumn{1}{c}{$J=1$}\\
\hline
0 & 3 &  6.8549 &  3.8100 & -0.1054 &&  17.8049 && {\bf -18.0662} & -18.6888 &  16.9358 && {\bf-27.9358} & -22.2457 & -16.7916 & -7.4441 &  {\bf14.0003} \\
  &   & -6.8564 & -3.8111 &  0.1047 && -17.8098 && {\bf18.0652} &  18.6893 & -16.9470 &&  {\bf27.9358} &  22.2423 &  16.7889 &  7.4388 & {\bf-14.0137} \\
\hline
9 & 3 &  6.1161 &  2.9767 & -0.9929 &&  14.5500 && {\bf-18.0820} & -16.3205 &  15.6554 && {\bf-27.9391} & -21.5026 & -15.0863 & -5.2879 &  {\bf13.9928} \\
  &   & -6.1179 & -2.9776 &  0.9928 && -14.5592 && {\bf 18.0807} &  16.3231 & -15.6716 &&  {\bf27.9391} &  21.4959 &  15.0801 &  5.2771 & {\bf-14.0167} \\
\hline\hline
\end{tabular}
}
\end{center}
\caption{Zeeman shift of the magnetic sublevels $M_J = -J$ (upper line) and $M_J = J$ (lower line) in a 0.02 G field (in kHz), for all hyperfine sublevels $(F,S,J)$ of the rovibrational states involved in the two-photon transition under study. The most Zeeman-insensitive transitions are highlighted by bold characters.}\label{TableZeeman}
\end{table}

\twocolumngrid

It appears that some of the hyperfine components connecting homologous spin states (i.e. states with the same $(F,S,J)$) benefit from a strong cancellation of Zeeman shifts. This occurs for $(F,S,J)=(1,1,4)$, $(1,2,5)$ and $(1,2,1)$ where the Zeeman splitting is respectively of 31.3, 6.6, and 4.5 Hz at 0.02~G. In the last two cases, the Zeeman structure is hidden within the natural linewidth of the transition and will not limit the resolution in any way. The most favorable component is $(F,S,J)=(1,2,5)$ since this hyperfine level has the highest population, allowing to get a stronger REMPD signal. There is only one dipole-allowed intermediate level for the two-photon transition, namely the $(v=4,L=2)$, $(F,S,J)=(1,2,4)$ level, so that the three-level approximation introduced in Sec.~II is well-justified in this case.

\section{Trapped ion dynamics}\label{ap_ion_dynamics}

In order to get a realistic description of the sympathetically cooled HD$^+$ ion velocities, we use a home-made simulation code taking into account the time-dependent trapping force, the Coulomb interaction and the laser cooling process (recoil due to absorption and emission of individual photons)~\cite{Schiller2003,Wubbena2012}. The laser-cooled ions are described as two-level atomic systems with a transition width $\Gamma_{\rm Be^+}=19.4$~MHz.

We assume a perfect linear quadrupolar Paul trap geometry with $r_0=3.5$~mm. The RF frequency $\Omega_{RF}$ is $2\pi\times 13.3$~MHz and the RF voltage amplitude is $V_0 =270$~V. The stability parameter for the radial confinement is $q=0.2$ for HD$^+$ and 0.067 for $^9$Be$^+$. A harmonic axial static potential provides axial confinement, with a trap frequency $\omega_z/2\pi =100$~kHz for Be$^+$ ions and 173~kHz for HD$^+$ ions. The Coulomb interaction between the ions, which is responsible for the sympathetic cooling, is taken into account without any approximations.

The Newton equations of motion are numerically integrated using a fixed step leap-frog algorithm~\cite{Verlet1967}. The time step $\delta t=2\times$~10$^{-10}$~s is chosen short enough to well represent the RF field, Coulomb collisions and laser absorption/emission cycles and to get converged results for simulation times up to 20~ms.

The laser interaction is described in terms of absorption, spontaneous or stimulated emission processes, thus including saturation effects. The laser beam has a wavelength $\lambda = 313.13$~nm and a TEM$_{00}$ Gaussian profile with a waist $w_0=1$~mm much larger than the ion cloud size. It is assumed to be perfectly aligned with the trap axis. The laser intensity $I$ and laser detuning $\delta_L$ are chosen close to optimal cooling conditions ($\delta_L = -\Gamma_{\rm Be^+}$, and $I = I_{\rm sat}/2$ where $I_{\rm sat}$ is the saturation intensity). At each time step, and for each laser-cooled ion in the ground state, the absorption probability is evaluated at the ion location and compared to a uniform random number generator between 0 and 1. In case an absorption occurs, the ion velocity is altered by a kick $\hbar{\bf k}/m$ where ${\bf k}$ is the photon wave vector. For laser-cooled ions in the excited state, the spontaneous (stimulated) emission is treated in a similar way but with a $\hbar k/m$ velocity kick with a uniformly randomized direction (a $-\hbar{\bf k}/m$ velocity kick)~\cite{Wubbena2012,Cagnac}.

A simulation is run in the following way. Ion position and velocities are randomized in a cylindrical volume around the trap center with a temperature $T\approx 10$~K. During the first 0.2~ms of the simulation, a huge drag force is applied to reach the Coulomb crystal regime where each ion oscillates around an equilibrium position. Then, the laser interaction is turned on and the ion cloud relaxes to its equilibrium temperature which is usually reached after 0.8~ms. Ion positions and velocities, mean secular kinetic energies, potential and Coulomb energies are periodically stored with a period of $4\times 10^{-8}$~s. With a pure sample of laser-cooled ions, we have checked that the ion cloud equilibrium temperature corresponds to the Doppler limit $k_BT=\hbar\Gamma_{Be^+}/2$ in the optimal cooling conditions.

\begin{figure}%[!h]
\includegraphics[width=7cm]{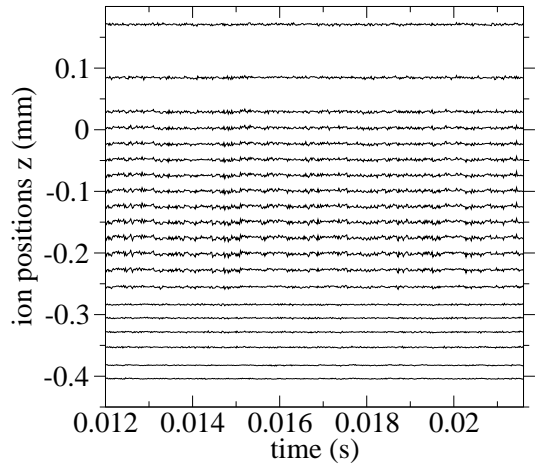}
  \caption{Typical axial trajectories around their equilibrium position for 20 sympathetically cooled HD$^+$ ions in a 400~Be$^+$ ions Coulomb crystal. Laser cooling conditions: detuning $\delta_L = -\Gamma_{\rm Be^+}$, saturation parameter $I/I_{\rm sat} = 1.5$.}
  \label{fig12_traj_z}
\end{figure}

\begin{figure}%[!h]
\includegraphics[width=8cm]{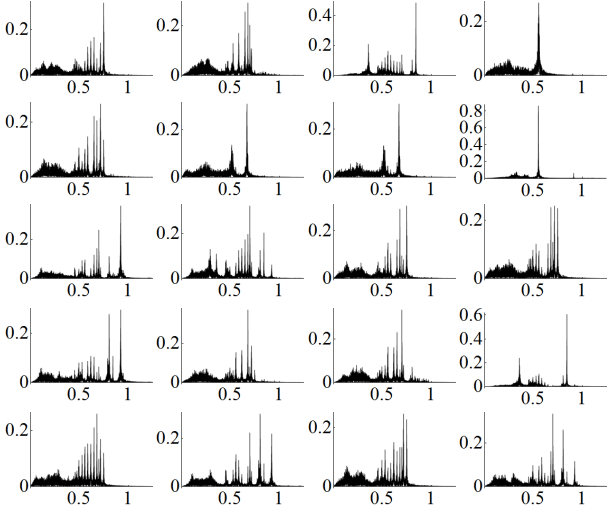}
  \caption{Spectrum of the velocity amplitudes obtained by FFT from the trajectories shown in Fig.~\ref{fig12_traj_z}. Horizontal scale is in MHz, vertical scale is in m/s/$\sqrt{\text{Hz}}$.} \label{fig13_fft_vz}
\end{figure}

Figure~\ref{fig12_traj_z} shows typical axial ($z$-axis) trajectories for 20~HD$^+$ ions that are sympathetically cooled by 400~Be$^+$ ions. The ions are nearly equally spaced and shifted in the direction of the incoming Be$^+$ cooling laser. The axial motion amplitude is in the $\mu$m range and the maximum axial velocities are of the order of 5~m/s. This gives a maximum Doppler effect $v/\lambda \approx 3.5$~MHz at the wavelength of the two-photon excitation lasers $\lambda = 1.44$~$\mu$m. The Doppler shift is larger than the oscillation frequencies, indicating that in the ion rest frame, the ions see motional sidebands with high modulation indexes. Figure~\ref{fig13_fft_vz} shows the axial velocity spectrum for each ion. Depending on the ion position within the cloud, the ion motion can be close to a pure harmonic motion or have a complex spectrum. This explains why the REMPD signal has to be averaged over the different ion trajectories.

%\vspace{1cm}

\section{Two-photon transition probability}\label{ap_2ph_proba}

We here consider a trapped particle with a three-level internal structure, undergoing one-dimensional harmonic motion at frequency $\Omega_{\rm vibr}$. The external degree of freedom is described quantum-mechanically and labelled by the vibrational quantum number $n$. Following time-dependent second order perturbation theory, the two-photon transition rate between states $|1,n_1\rangle$ and $|3,n_3\rangle$ is given by
\begin{eqnarray}
\Gamma_{\rm 2ph} &=&  \left | \sum_{n_2 = 0}^\infty \frac{\Omega_{12}\Omega_{23} <n_3| e^{-ik'z} |n_2><n_2| e^{ikz} |n_1>}{(\delta_{12} - i\frac{\Gamma_2}{2} + (n_1 - n_2) \Omega_{\rm vibr})} \right |^2 \nonumber \\
&&\times \frac{\Gamma_{3}^{\rm eff} }{(\delta_{13}+ (n_1 - n_3) \Omega_{\rm vibr})^2 + \frac{(\Gamma_3^{\rm eff})^2}{4}}.
\label{eq_two-photon_proba_general}
\end{eqnarray}
Assuming the detuning $\delta_{12}$ is much larger than both the intermediate level width and the vibration frequency, and summing over $n_2$, the first term in Eq.~(\ref{eq_two-photon_proba_general}) can be simplified leading to
\begin{eqnarray}
\Gamma_{\rm 2ph} &=&  \frac{ \left | \Omega_{12}\Omega_{23} <n_3| e^{i\delta\!k\ z} |n_1>\right |^2}{\delta_{12}^2}
\nonumber \\
&&\times \frac{\Gamma_{3}^{\rm eff} }{(\delta_{13} + (n_1 - n_3) \Omega_{\rm vibr})^2 + \frac{(\Gamma_3^{\rm eff})^2}{4}}.
\label{eq_proba_2ph_0}
\end{eqnarray}
The denominator of the second factor shows that the two-photon transition probability exhibits sidebands separated by $\Omega_{\rm vibr}$. The amplitudes of the sidebands are given by the matrix element $<n_3| e^{i\delta\!k\ z} |n_1>$~\cite{Leibfried2003,Wineland1979,Cahill1969} with
\begin{equation}
\left|<n+s| e^{i\eta(a+a^\dag)} |n>\right|=e^{-\eta^2/2}\eta^{|s|}\sqrt{\frac{n_<!}{n>!}}L_{n_<}^{|s|}(\eta^2),
\label{eq_matrix_elem_LambDicke}
\end{equation}
where $n_<$ and $n_>$ are the lesser and greater of $n$ and $n+s$, and $\eta=\delta\!k\sqrt{\hbar/(2\ m\Omega_{\rm vibr})}$. $L^{s}_{n}$ are the generalized Laguerre polynomials, and $a$, $a^\dag$ are the creation and annihilation operators associated with the harmonic confinement. In the Lamb-Dicke regime where the oscillation amplitude is much smaller than the effective wavelength $2\pi/\delta k$,  this matrix element is $\approx \delta_{n_1,n_3}$ and the two-photon rate further simplifies to
\begin{equation}\label{eq_proba_2ph}
\Gamma_{\rm 2ph} =   \frac{ \Omega_{12}^2\Omega_{23}^2}{\delta_{12}^2}\frac{\Gamma_{3}^{\rm eff} }{\delta_{13}^2 + \frac{(\Gamma_3^{\rm eff})^2}{4}}.
\end{equation}
On two-photon resonance where $\delta_{13}$~=~0, it is given by
\begin{equation}
\Gamma_{\rm 2ph} =   \frac{ \Omega_{12}^2\Omega_{23}^2}{\delta_{12}^2}\frac{4}{\Gamma_{3}^{\rm eff} }.\label{eq_proba_2ph_simple}
\end{equation}

\section{Laser phase noise simulation}\label{ap_laser_phase_noise}

In this Appendix, we describe the phase noise generator we have implemented to simulate the laser Lorentzian lineshape. Let $f(t)$ denote the instantaneous laser frequency, and $\delta f$ the laser frequency noise. It is linked to the laser phase noise by $\delta f=\frac{1}{2\pi}\frac{d\varphi(t)}{dt}$. Laser phase noise $\varphi(t)$ is usually depicted as a centered stationary Gaussian process with a white frequency noise (single sided) spectral density $S_{\delta f}(\omega)$ in a bandwith 2$\pi$B~\cite{Elliott1982}. The variance of the laser frequency noise is given by $\langle(\delta f)^2\rangle=BS_{\delta f}(\omega)$. The laser linewidth $\Delta_{\rm FWHM}$ is defined by the full width at half maximum of the hypothetical beat note spectrum of the laser with a perfect noiseless laser. It can be expressed in an integral form as a function of $S_{\delta f}(\omega)$~\cite{Elliott1982}. If $\langle(\delta f)^2\rangle \ll B^2$, the lineshape is Lorentzian with $\Delta_{\rm FWHM}=\pi S_{\delta f}(\omega)$. If $\langle(\delta f)^2\rangle \gg B^2$, the lineshape is Gaussian with $\Delta_{\rm FWHM}=2\sqrt{2\ln2}\sqrt{S_{\delta f}B}$. For intermediate cases, the linewidth was evaluated by numerical computation of an integral leading to an empirical interpolating formula~\cite{DiDomenico2010}
\begin{equation}
\Delta_{\rm FWHM}  = S_{\delta f}\frac{\sqrt{8\ln2\ B/S_{\delta f}}}{\left(1+\frac{8\ln2}{\pi^2}\frac{B}{S_{\delta f}} \right)^{1/4}}\label{eq_noise_empiric}.
\end{equation}

The frequency noise and phase noise spectral densities are linked by $S_{\delta f}(\omega)=\left(\frac{\omega}{2\pi}\right)^2 S_{\varphi}(\omega)$ so a white frequency noise in a bandwidth $B$ corresponds to a $1/\omega^2$ phase noise spectral density with $0<\omega\leq2\pi B$. The Wiener-Khintchin theorem states that $S_{\varphi}(\omega)=|\tilde{\varphi}(\omega)|^2$ where $\tilde{\varphi}$ is the Fourier transform of $\varphi(t)$. Therefore, the desired laser phase noise can be obtained by randomly generating the Fourier components $\tilde{\varphi}(\omega)$ and performing an inverse fast Fourier transform.

The discretization is done in the following way. The simulation duration $T$ and the integration time step $\delta t$ sets the number $N=T/\delta t$ of $\varphi$ values $\varphi_j=\varphi(j\delta t)$.
It also sets the maximum Fourier frequency $f_{\rm max}=1/2\delta_t$ and the frequency resolution $1/T$. The corresponding Fourier frequencies and discretized Fourier components are $\omega_j=2\pi j/T$ and $\tilde{\varphi}_j$ with $-N/2\leq j\leq N/2$. The maximum Fourier frequency has to be larger than the noise bandwidth, i.e. $B<\delta t/2$. The phase noise discretized Fourier components are randomly generated following $\tilde{\varphi}(\omega_j)=\frac{K}{\omega_j} e^{i\phi_j}$ for $0<j\leq BT$ and set to 0 for $j=0$ and $BT<j\leq N/2$. Since the phase noise is a real process, the negative frequency Fourier components are equal to the positive ones hence $\tilde{\varphi}_{-j}=\tilde{\varphi}_j^*$. To generate a random phase noise, the complex argument of the Fourier components $\phi_j$ is uniformly randomized between 0 and 2$\pi$. The noise level $K$ is linked to the variance of the laser frequency noise by $K=\sqrt{S_{\delta f}f_{\rm max}}$. Finally, the FFT of the phase noise components is computed using the fftw3 FORTRAN subroutine library to obtain the time-dependent phase noise that is used by the OBE numerical solver.

\begin{figure}%[!h]
\includegraphics[width=8cm]{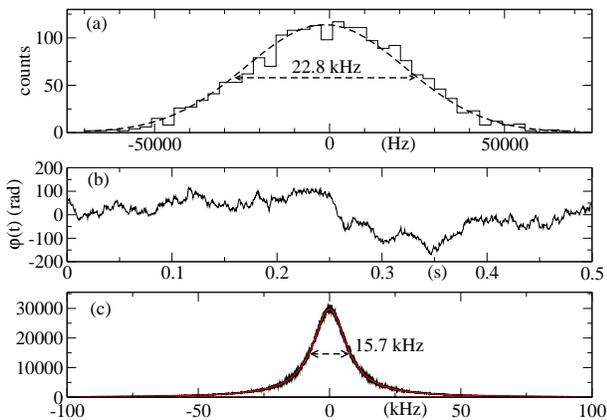}
  \caption{(a) Histogram of the instantaneous frequency $f(t)$ at time $t=0$ for 2000 realizations of the phase noise. (b) a single realization of $\varphi(t)$. (c) Black: Averaged laser lineshape for the 2000 phase noise realizations. Red: Lorentzian fit.}
  \label{fig15_phase_noise}
\end{figure}

\begin{figure}%[!h]
\includegraphics[angle=-90,width=8cm]{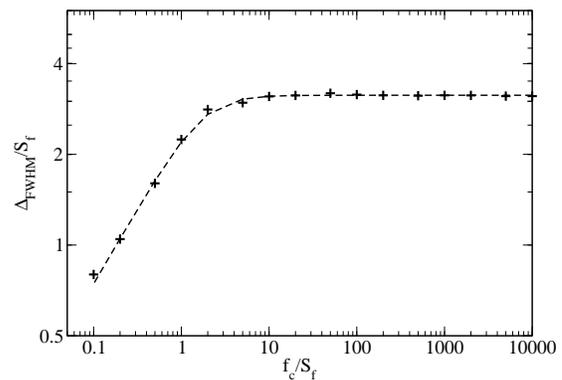}
  \caption{Comparison of the laser FWHM (crosses) with the empirical formula (dashed line) given in Eq.~(\ref{eq_noise_empiric}).}
  \label{fig16_FWHM_test}
\end{figure}

Figure~\ref{fig15_phase_noise}(a) shows the histogram of the instantaneous frequency $f(t)$ obtained for 2000 realizations of the noise process with $T=0.5$~s, $\delta t=4\times\ 10^{-7}$~s, $S_{\delta f}=5000$~Hz$^2$/Hz and $B=100$~kHz. A Gaussian fit gives a 22.8~kHz standard deviation in agreement with $\sqrt{S_{\delta f}B}=22.4$~kHz. Figure~\ref{fig15_phase_noise}(b) shows a realization of $\varphi(t)$ and Fig.~\ref{fig15_phase_noise}(c) shows the average lineshape of the beatnote. The Lorentzian width is 15.7~kHz in perfect agreement with $\pi S_{\delta f}$. We have varied the frequency noise spectral density $S_{\delta\!f}$ from 10 to 10$^6$~Hz$^2$/Hz and determined the FWHM of the line. Figure~\ref{fig16_FWHM_test} shows that it follows the empirical formula and thus the expected linewidth behavior.

Finally, to generate laser phase noise with a Lorentzian lineshape, one has to fulfill the conditions $B\gg 6S_{\delta f}$ and choose $S_{\delta f}=\Delta_{\rm FWHM}/\pi$, so the noise bandwidth must obey $B\gg 6/\pi\Delta_{\rm FWHM}$.


\begin{acknowledgements}
V.Q.T. would like to acknowledge Ile de France region for his PhD grant. J.C.J.K. would like to acknowledge the Netherlands Organisation for Scientific Research (NWO) and the Dutch Foundation for Fundamental
Research on Matter (FOM) for support.
The Authors acknowledge the COST action MP1001 {\it Ion Traps for Tomorrow Applications} (IOTA).
\end{acknowledgements}

\end{document}